\documentclass[preprint]{aastex}

\usepackage{amsmath}
\usepackage{subfigure}
\usepackage{booktabs}
\usepackage{ulem}
\usepackage{xcolor}
\newcommand{\saltrms}{\ensuremath{\mathrm{0.148\pm0.015}}}

\shorttitle{Twin Type~Ia Supernovae}
\shortauthors{Fakhouri et al.}

\begin{document}

\title{
    Improving Cosmological Distance Measurements \\ Using Twin Type~Ia Supernovae
    }

\author
{
    H.~K.~Fakhouri,\altaffilmark{1,2}
    K.~Boone,\altaffilmark{1,2}
    G.~Aldering,\altaffilmark{1}
    P.~Antilogus,\altaffilmark{3}
    C.~Aragon,\altaffilmark{1}
    S.~Bailey,\altaffilmark{1}
    C.~Baltay,\altaffilmark{4}
    K.~Barbary,\altaffilmark{2}
    D.~Baugh,\altaffilmark{5}
    S.~Bongard,\altaffilmark{3}
    C.~Buton,\altaffilmark{6}
    J.~Chen,\altaffilmark{5}
    M.~Childress,\altaffilmark{7} 
    N.~Chotard,\altaffilmark{6}
    Y.~Copin,\altaffilmark{6}
    P.~Fagrelius,\altaffilmark{1,2}
    U.~Feindt,\altaffilmark{8}
    M.~Fleury,\altaffilmark{3}
    D.~Fouchez,\altaffilmark{9}
    E.~Gangler,\altaffilmark{10}  
    B.~Hayden,\altaffilmark{1}
    A.~G.~Kim,\altaffilmark{1}
    M.~Kowalski,\altaffilmark{8,11}
    P.-F.~Leget,\altaffilmark{10}
    S.~Lombardo,\altaffilmark{8}
    J.~Nordin,\altaffilmark{1,8}
    R.~Pain,\altaffilmark{3} \\
    E.~Pecontal,\altaffilmark{12}
    R.~Pereira,\altaffilmark{6}
    S.~Perlmutter,\altaffilmark{1,2}
    D.~Rabinowitz,\altaffilmark{4}
    J.~Ren,\altaffilmark{1,2}
    M.~Rigault,\altaffilmark{8} \\
    D.~Rubin,\altaffilmark{1,13}
    K.~Runge,\altaffilmark{1}
    C.~Saunders,\altaffilmark{1,2}
    R.~Scalzo,\altaffilmark{7}
    G.~Smadja,\altaffilmark{6} 
    C.~Sofiatti,\altaffilmark{1,2} \\
    M.~Strovink,\altaffilmark{1,2}
    N.~Suzuki,\altaffilmark{1}
    C.~Tao,\altaffilmark{5,9}
    R.~C.~Thomas,\altaffilmark{14}
    B.~A.~Weaver\altaffilmark{15} \\
    (The Nearby Supernova Factory)
}
\altaffiltext{1}
{
    Physics Division, Lawrence Berkeley National Laboratory, 
    1 Cyclotron Road, Berkeley, CA, 94720
}
\altaffiltext{2}
{
    Department of Physics, University of California Berkeley,
    366 LeConte Hall MC 7300, Berkeley, CA, 94720-7300
}
\altaffiltext{3}
{
    Laboratoire de Physique Nucl\'eaire et des Hautes \'Energies,
    Universit\'e Pierre et Marie Curie Paris 6, Universit\'e Paris Diderot Paris 7, CNRS-IN2P3, 
    4 place Jussieu, 75252 Paris Cedex 05, France
}
\altaffiltext{4}
{
    Department of Physics, Yale University, 
    New Haven, CT, 06250-8121
}
\altaffiltext{5}
{
    Tsinghua Center for Astrophysics, Tsinghua University, Beijing 100084, China 
}
\altaffiltext{6}
{
    Universit\'e de Lyon, F-69622, Lyon, France ; Universit\'e de Lyon 1, Villeurbanne ; 
    CNRS/IN2P3, Institut de Physique Nucl\'eaire de Lyon.
}
\altaffiltext{7}
{
    Research School of Astronomy and Astrophysics,
    Australian National University,
    Canberra, ACT 2611, Australia
}
\altaffiltext{8}
{
    Institut fur Physik,  Humboldt-Universitat zu Berlin,
    Newtonstr. 15, 12489 Berlin
}
\altaffiltext{9}
{
    Centre de Physique des Particules de Marseille, 
    Aix-Marseille Universit\'e , CNRS/IN2P3, 
    163 avenue de Luminy - Case 902 - 13288 Marseille Cedex 09, France
}
\altaffiltext{10}
{
    Clermont Universit\'e, Universit\'e Blaise Pascal, CNRS/IN2P3, Laboratoire de Physique Corpusculaire,
    BP 10448, F-63000 Clermont-Ferrand, France
}
\altaffiltext{11}
{
    DESY, D-15735 Zeuthen, Germany
}
\altaffiltext{12}
{
    Centre de Recherche Astronomique de Lyon, Universit\'e Lyon 1,
    9 Avenue Charles Andr\'e, 69561 Saint Genis Laval Cedex, France
}
\altaffiltext{13}
{
    Department of Physics, Florida State University,
    315 Keen Building, Tallahassee, FL 32306-4350
}
\altaffiltext{14}
{
    Computational Cosmology Center, Computational Research Division, Lawrence Berkeley National Laboratory, 
    1 Cyclotron Road MS 50B-4206, Berkeley, CA, 94720
}
\altaffiltext{15}
{
    Center for Cosmology and Particle Physics,
    New York University,
    4 Washington Place, New York, NY 10003, USA
}

\begin{abstract}

We introduce a method for identifying ``twin'' Type Ia supernovae, and using
them to improve distance measurements. This novel approach to Type~Ia supernova
standardization is made possible by spectrophotometric time series observations
from the Nearby Supernova Factory (SNfactory). We begin with a well-measured
set of supernovae, find pairs whose spectra match well across the entire
optical window, and then test whether this leads to a smaller dispersion in
their absolute brightnesses. This analysis is completed in a blinded fashion,
ensuring that decisions made in implementing the method do not inadvertently
bias the result. We find that pairs of supernovae with more closely matched
spectra indeed have reduced brightness dispersion. We are able to standardize
this initial set of SNfactory supernovae to $0.083\pm0.012$~magnitudes,
implying a dispersion of $0.072\pm0.010$~magnitudes in the absence of peculiar
velocities. We estimate that with larger numbers of comparison SNe, e.g, using
the final SNfactory spectrophotometric dataset as a reference, this method will
be capable of standardizing high-redshift supernovae to within 0.06 --
0.07~magnitudes. These results imply that at least 3/4 of the
    variance in Hubble residuals in current supernova cosmology analyses
    is due to previously unaccounted-for astrophysical differences among the
supernovae.

\end{abstract}

\keywords{cosmology: observations -- supernovae: general}

\section{Introduction}

Type~Ia supernovae (SNe~Ia) have a well-established history as
standardized candles for  cosmological distance measurements. In
the late 1990s, studies of high-redshift Type~Ia supernovae led to
the discovery of the accelerated expansion of the universe -- a
remarkable and paradigm-shifting discovery \citep{riess98a,
perlmutter99a}. Before standardization is applied, among ``normal''
SNe~Ia there exists a dispersion of approximately 40\% \citep{kim97}.
The physical source of this variation is not fully understood, but
a number of methods have been developed to reduce this natural
dispersion \citep{vaughan95,riess96,tripp98,lwang03,xwang05,
bailey09,mandel11,barone12,kim13}. The most established approach
uses the light curve width and color as standardization parameters.
SNe~Ia with broader light curves tend to be brighter \citep{phillips93}.
To some degree, the color correction seems able to account for
extrinsic reddening due to dust and intrinsic color differences
(presumably arising from differences in the physics of the supernova
explosion). With these corrections, the variation in brightness
drops to $\sim15$\%, allowing precision cosmology measurements using
large samples of SNe~Ia.

Since the discovery of the accelerating expansion, many large Type~Ia
supernova programs have been undertaken, resulting in sufficiently
large data sets that statistical errors have become subdominant to
systematic errors \citep{sullivan11a, betoule14}. Much of the effort
in reducing systematic errors has focused on the flux calibration
of SNe~Ia at low-redshift relative to those at high redshift
since this error is relatively large and there are clear, though
technically challenging, means of addressing it \citep{juramy08,
stubbs10, conley11a, tonry12, betoule14, lombardo14, scolnic14,
rubin15}.

If the remaining dispersion of 15\% were 
due to a statistically homogeneous process, then
correction of flux calibration issues would clear the way for more
large SN surveys using conventional light curve width and color
standardization. However, there is ample evidence that the remaining
dispersion is not statistically homogeneous. Only relatively recently have significant
biases correlated with host galaxy environment been uncovered, and
the physical explanation for these observed correlations is still under debate 
\citep{kelly10,sullivan10,lampeitl10,childress13b,rigault13,
rigault15}. The size of these environmental biases is $\sim10$\%,
quite large relative to the standardized brightness dispersion, yet
correction for these biases still leaves a significant dispersion
(e.g., $\sim10$\% in \citet{rigault13,rigault15}) that might hide
additional astrophysical biases. The diversity of SNe~Ia spectral
features, which have been ascribed to subpopulations of otherwise
normal SNe~Ia, also contributes significant systematic errors that
are also reflected in the standardize brightness dispersion
\citep{xwang09, fk11,chotard11,saunders15}. Dust correction could
also introduce systematic errors. There is not only evidence for a
variety in the extinction curves of the dust itself \citep[][and
references therein]{patat15}, but also for a strong interplay between
SN spectral features and the lightcurve ``color'' appropriate for
applying extinction corrections \citep{chotard11,fk11}.

Suppression of such astrophysical systematics requires separation
of intrinsic supernova explosion differences from those of dust.
However, disentangling the two has proven a major challenge
due to the lack of a clearly identifiable set of observational indicators.  
Thus, while decreasing the dispersion in standardized brightnesses 
has a clear statistical payoff, its more important impact is in 
the {\it implicit} suppression of systematic errors.

Techniques to decrease the brightness dispersion continue to be
developed. Various spectral metrics have been tried, mostly using
spectra near maximum light. For instance, \citet{bailey09} were
able to achieve a $0.128\pm0.011$~mag dispersion using a one-parameter
spectral correction. This is notable in that this one parameter
seems to have captured much of the variation from dust, intrinsic
color, and whatever physics is encoded in the lightcurve width
(likely total mass; \citet{scalzo14}). Another approach has been to
attempt to evade dust extinction altogether and to therefore focus
on restframe
NIR wavelengths. Theoretical modeling suggests that the restframe
NIR also may have less intrinsic brightness variation \citep{kasen06}.
Recent filter-photometry results demonstrate standardization as
good as $0.085\pm0.016$~mag for nearby SNe~Ia (\citet{barone12},
\citet{mandel11}). These types of studies still do not resolve the
fact that some of the observed color arises from dust while some
is intrinsic color from the specifics of the explosion.

Using new more detailed measurements of Type~Ia supernova time series
obtained by the Nearby Supernova Factory
\citep[SNfactory;][]{aldering02},
we are able to go further in exploring standardization techniques using spectra
across all phases. A time series of spectrophotometric
observations offers a look at the physics of the explosion through
the expanding photosphere.  Each phase observation is a snapshot
of the elements visible at that stage of the expansion, and their 
absorption-weighted velocities.  If two
time series were to match in flux at all wavelengths and times, they
would presumably represent the same physical explosion process. Use
of such differential techniques is a well-established experimental
method for reducing systematics and increasing sensitivity. We call
two such time series ``twin supernovae''.

A distinct advantage of this twin supernova methodology is the
ability to separately account for color arising from dust extinction.
The relative amount of dust extinction affecting
two spectrophotometric time series can be fit and removed. Any 
remaining color difference can be ascribed to intrinsic color 
differences (minus any intrinsic color behavior that happens to
exactly mimic the dust color-extinction relation). Two
objects that are true twins would have the same intrinsic
color and the same spectral features at all times. For pairs of
supernovae that are twins, we expect a very small brightness difference,
as they represent the same explosion physics.  If this is indeed the case,
cosmology analyses could treat twin SNe~Ia as true standard candles, with no
lightcurve width or intrinsic color corrections necessary.

For this analysis, we use spectrophotometric time series with
host-galaxy subtractions that are unique to the SNfactory. Much of
this dataset is from untargeted searches in the Hubble flow, and
thus not only mimics the subtypes of Type~Ia supernovae that are
found in high-redshift searches, but also is in the redshift regime
where host-galaxy peculiar velocity uncertainties are substantially
less than the existing dispersion in standardized SNe~Ia 
brightnesses. Ideally, a dataset designed for a twin
supernova analysis would consist of frequent observations, to ensure
that the analysis captures relevant spectral feature changes.  The 
SNfactory dataset typically has spectrophotometric
observations every 2 to 3 nights.  Section~\ref{sec:data} describes
the SNfactory sample used for the twins analysis reported here.  Since
measurement of dispersion is highly dependent on the tails of the
residual distribution, we were careful to finalize the sample
selection criteria and analysis methodology before examining the
brightness scatter in order to remove the possibility that analysis
choices would artificially suppress the dispersion.

Section~\ref{sec:interp} then introduces the techniques we developed to
directly compare the supernovae observations to each other. The
analysis is discussed in Section~\ref{sec:rank}; we discuss both a
near-maximum analysis, and a weighted full time series analysis. In
Section~\ref{sec:deltam} the resulting dispersions for
each of these analyses is determined. In Section~\ref{sec:discussion}, we
compare the results to other standardization techniques, and 
discuss the application of the twins method to new high-redshift SNe~Ia,
before presenting our conclusions in Section~\ref{sec:conclusions}.

\section{Supernova Data Set}\label{sec:data}

Here we describe the data set used for this analysis. We 
include details of the observations, a short discussion 
of the time series interpolation method since it enters 
into the selection process, and the criteria and outcome 
of the final sample selection. Steps needed to prepare the
data for analysis are also described.

\subsection{Observations}\label{sec:obs}

The spectrophotometric time series used in this study were obtained by the Nearby Supernova Factory using the  SuperNova Integral Field Spectrograph \citep[SNIFS,][]{lantz04a}. SNIFS consists of a high-throughput wide-band pure-lenslet integral field spectrograph \citep[IFS, ``\`a la TIGER;''][]{bacon95a,bacon01a}, a
photometric channel to image the field in the vicinity of the IFS 
for atmospheric transmission monitoring simultaneous with spectroscopy, and an acquisition/guiding channel. The IFS possesses
a fully-filled $6.^{\prime\prime}4 \times 6.^{\prime\prime}4$ spectroscopic field of view subdivided into a grid of $15 \times 15$ spatial elements, a dual-channel spectrograph covering 
3200--5200~\AA\ and 5100--10000~\AA\ simultaneously, and an 
internal calibration unit (continuum and arc lamps).  SNIFS is
mounted on the south bent Cassegrain port of the University of 
Hawaii 2.2~m telescope on Mauna Kea, and is operated remotely.  

Observations primarily targeted SNe~Ia that, at the time, were 
expected to be at or before maximum light and in the redshift 
range $0.03<z<0.08$, with a nominal cadence of 2--3 nights. For
    spectra near maximum, the median signal-to-noise over all SNe and restframe
    wavelengths 3300--8600~\r{A} is $\sim42$ per
1000~km~s$^{-1}$ interval. By comparison, this signal-to-noise is $\sim18$
for spectra between 20--25 days after maximum. For a typical
near-maximum spectrum, this signal-to-noise varies from $\sim10$ near the
wavelength extremes
of the two spectrograph channels to $\sim50$ near their central wavelengths.
Multiple spectrophotometric standard stars were regularly observed each
night in order to
measure the instrument calibration and atmospheric extinction.
Parallel imaging was obtained so that it could be employed to remove
dimming due to clouds on nights that proved to be non-photometric.
Roughly a year or more after the nominal date of maximum, final reference
observations were obtained in order to subtract the host galaxy
and SN signal.

Spectra of all targets were reduced using the dedicated SNfactory data reduction pipeline, similar to that presented in \S~4
of \cite{bacon01a}. A brief discussion of the software pipeline is presented in \citet{aldering06a} and is updated in \citet{scalzo10a}. Detailed discussions of the flux calibration and host-galaxy subtraction are provided in \citet{buton13} and \citet{bongard11}, respectively.  
Other spectral and lightcurve results from these data have been presented in \citet{bailey09,chotard11,thomas11,scalzo12,childress13a,childress13b,feindt13,kim13,rigault13,kim14,scalzo14,saunders15}.

\subsection{Interpolation of Supernova Time Series}\label{sec:interp}

For the twins analysis it is necessary to compare candidate twin supernovae at the same phases. Realistic observing programs are unable to generate an observation of each SN on every night, so such a comparison requires an estimated spectrum 
using data at other phases. The quality of such an interpolation affects which spectral time series are appropriate for this study.
Here we briefly discuss the method we have implemented.

We have chosen to use Gaussian Process regression \citep{rasmussen05} to
predict SN time series at desired phases. To take advantage of all the available data, in comparing SN$_A$ to SN$_B$, we first predict SN$_A$ at the phases of SN$_B$ and then vice versa.  Each pair is given a ranking (discussed in Section~\ref{sec:nearmaxtwinness}) by combining the results of both directions.

A Gaussian Process (GP) gives a non-parametric reconstruction of a function from data. It requires the selection of a mean function, which is an initial estimate of the average supernova 
flux at all phases and wavelengths. We use the spectral template of \citet{hsiao} for our mean function. The end result is not highly dependent on the choice of this mean function, so long as a reasonable function is used, and the gaps in data are
not large compared to the typical length scale of the GP mean function. The Gaussian Process uses a kernel to specify the correlation lengths in wavelength and time. For the current analysis, the parameters of the kernel, called hyperparameters, are optimized separately for each SN time series. Once these hyperparameters are
found, both the predicted SN flux and an estimate of the flux uncertainty can be generated for any phase of interest. Appendix \ref{sec:gp_params} details the kernel and hyperparameters used in this analysis.

\begin{figure}
    \subfigure{
        \subfigure{
            \includegraphics[width=0.45\textwidth]{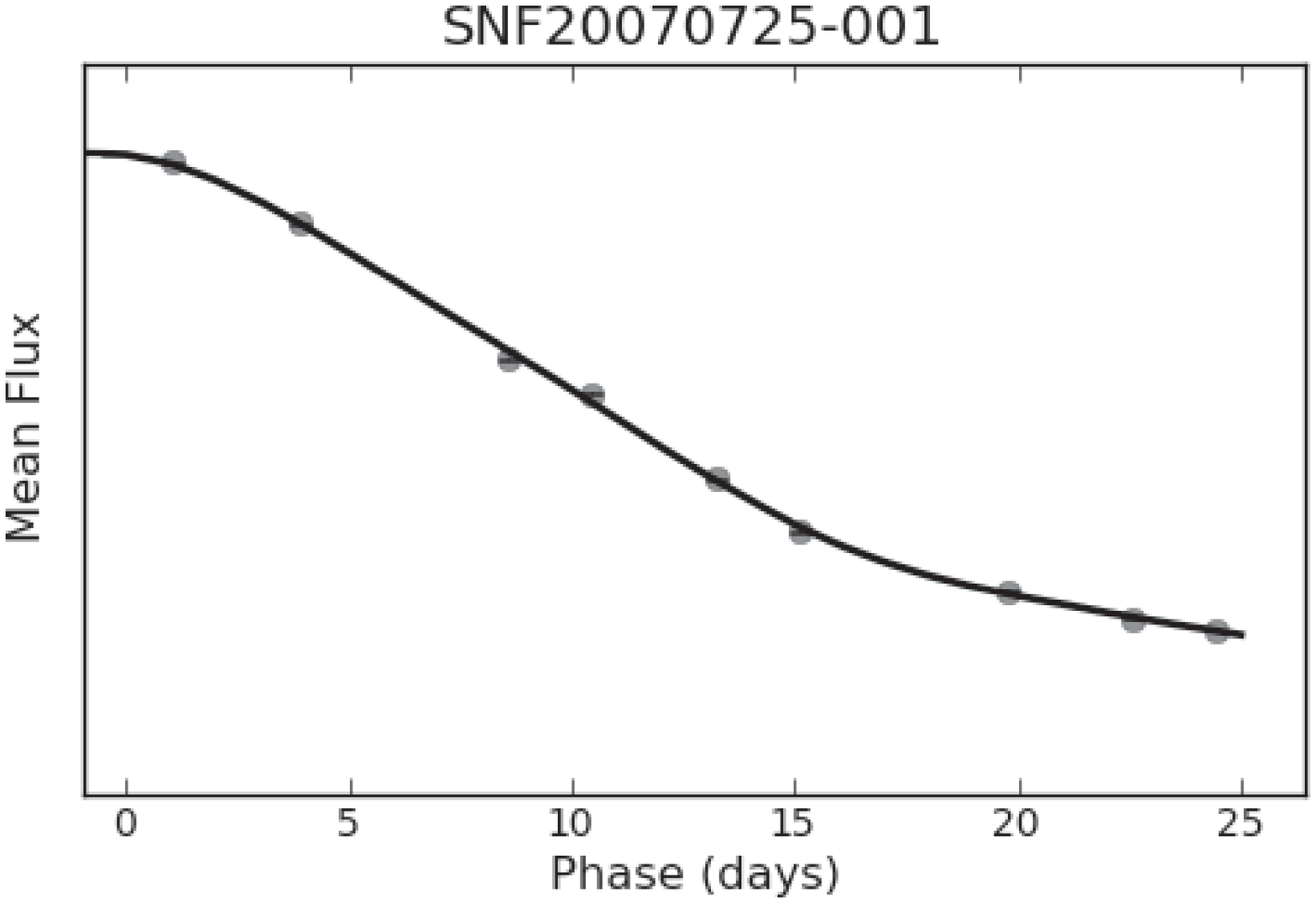}
        }
        \subfigure{
            \includegraphics[width=0.45\textwidth]{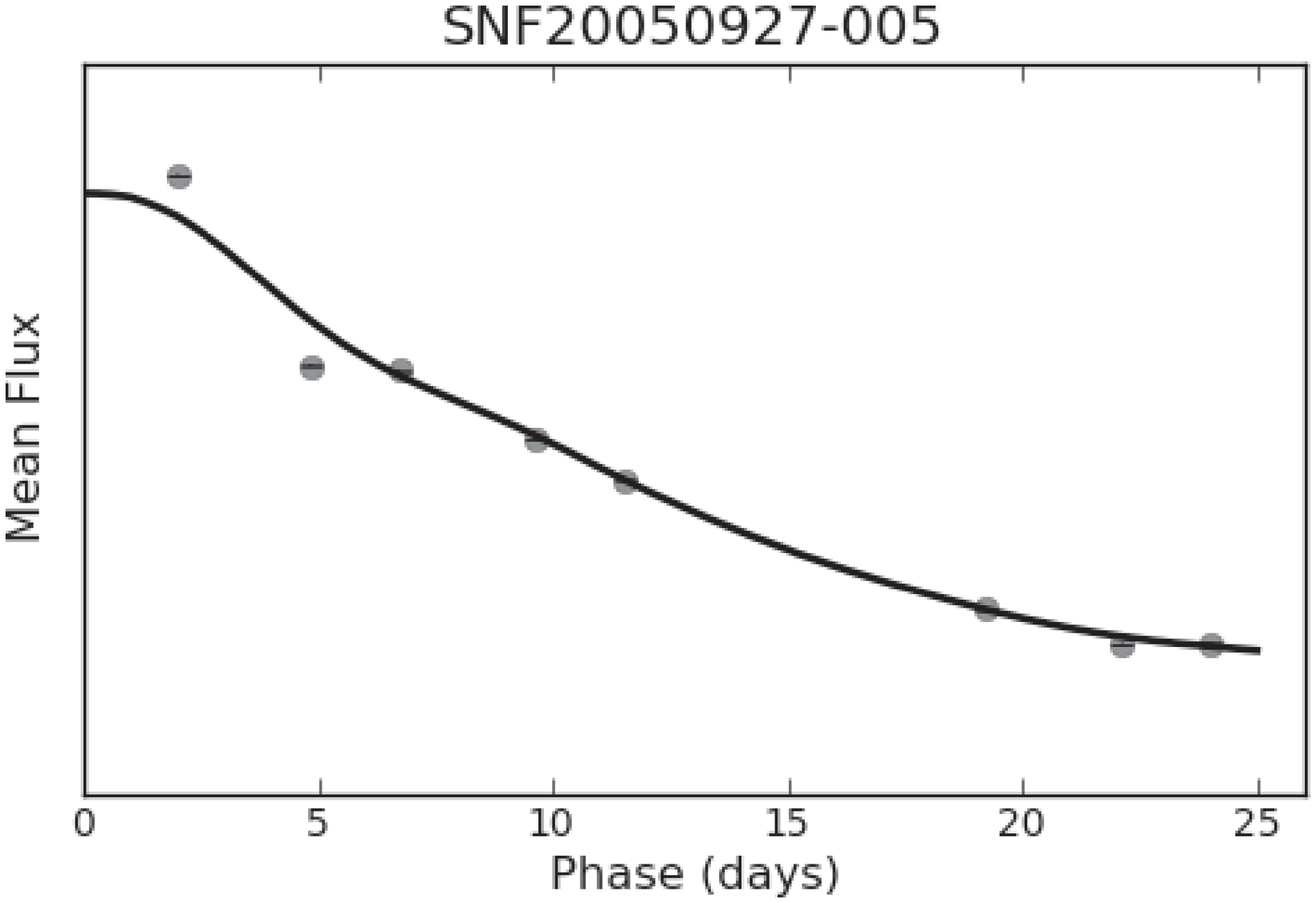}
        }
    }
    \subfigure{
    \subfigure{
    \includegraphics[width=0.45\textwidth]{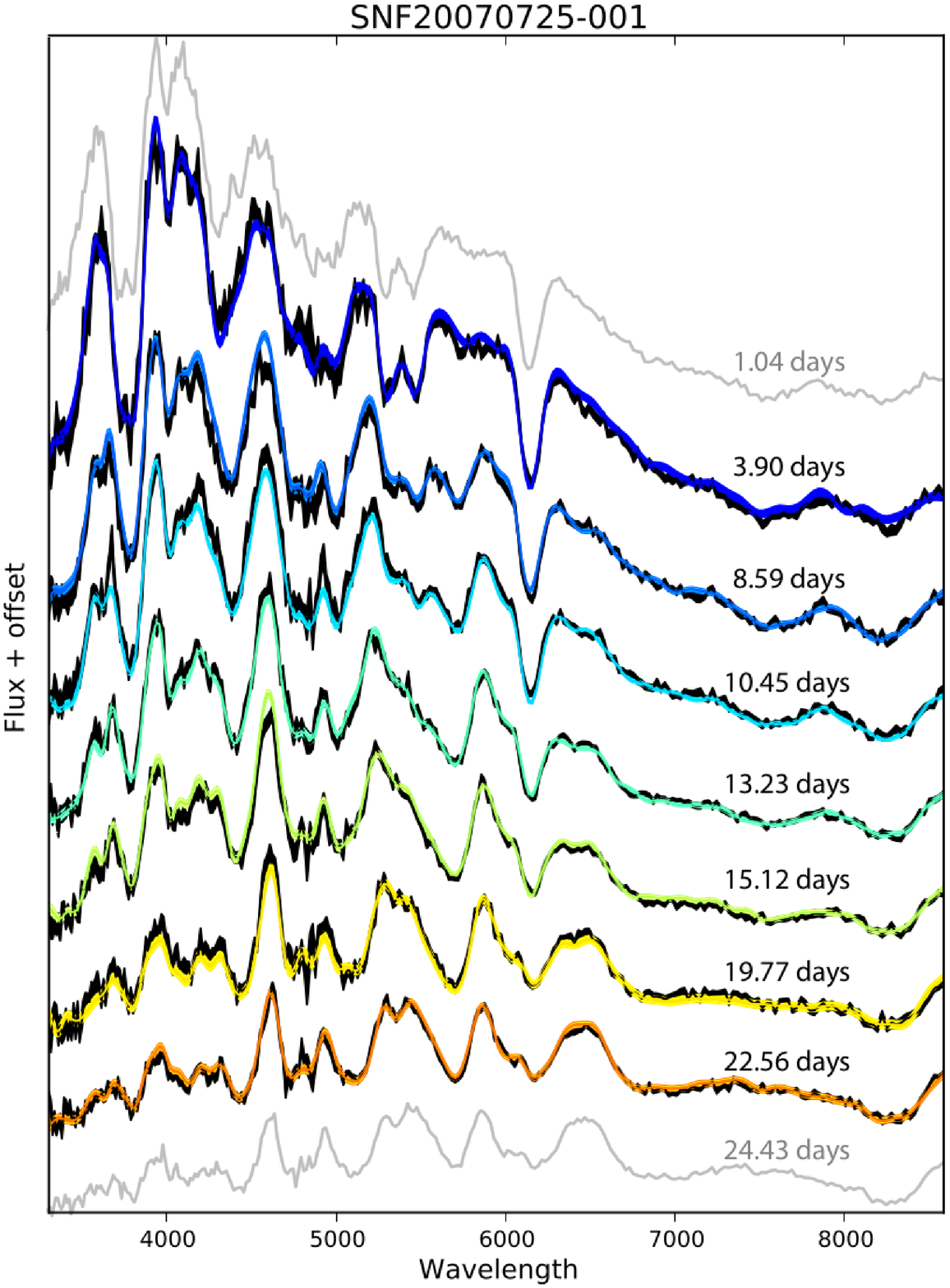}
}
    \subfigure{
    \includegraphics[width=0.45\textwidth]{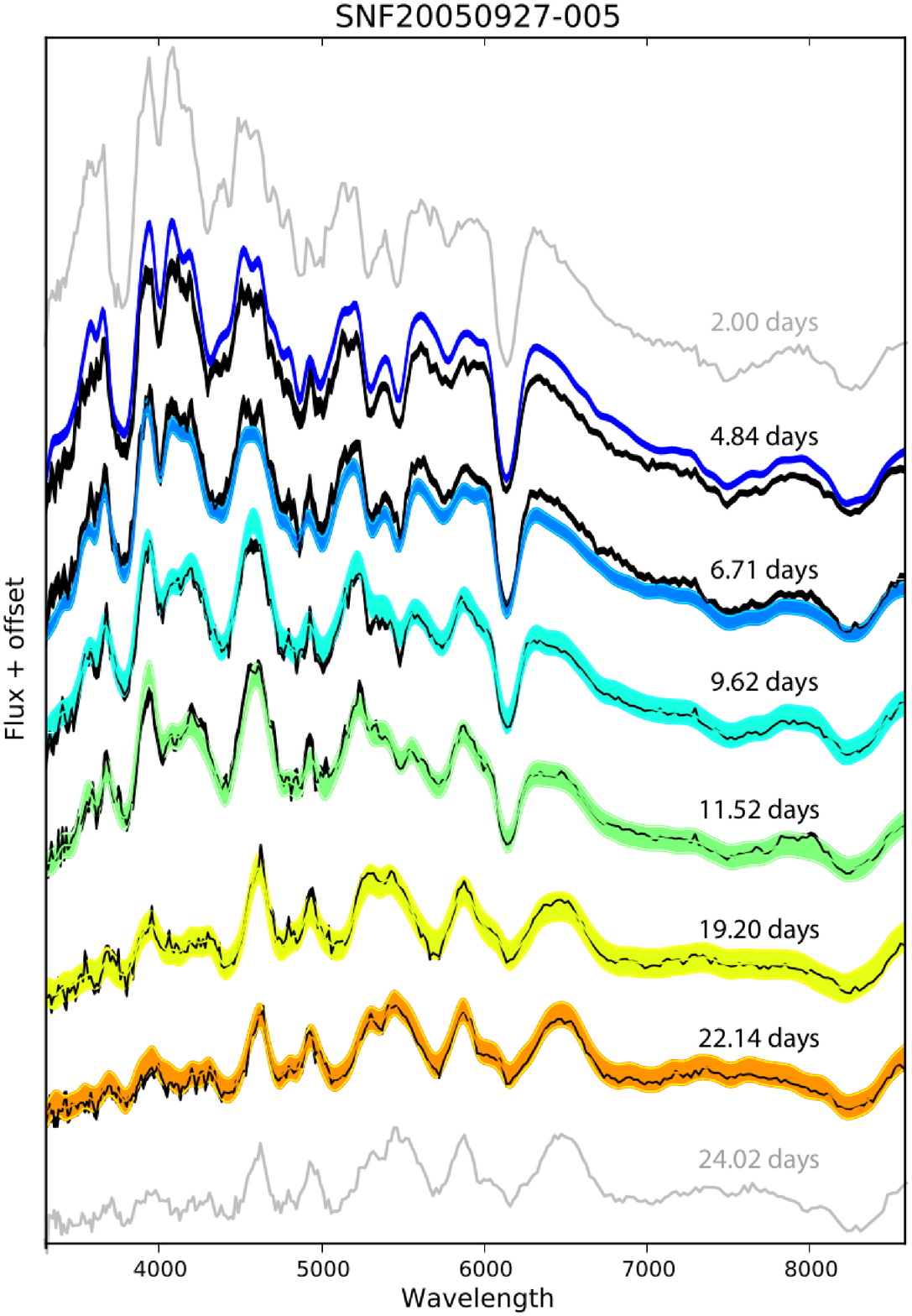}
}
}
    \caption{Wavelength-averaged Gaussian Process predictions for two
        supernovae. The top images show the wavelength averaged lightcurves. The bottom
        images show the results of the leave-one out test. The data for each phase are 
        shown in black, and the GP prediction for the leave-one-out test for that phase
        are shown in color. Left: SNF20070725-001, an example of a supernova that passes 
        the leave-one out test. Right: SNF20050927-005, an example of a supernova which 
        fails the leave-one out test. Several spectra are not reproduced well by the GP
        prediction of the data at other phases. To be conservative, we drop this supernova 
        while blinded. Note that the first and last data phases (shown in grey)
        are excluded  from the leave-one-out test as we do not expect the GP to be able to 
        extrapolate the endpoints well beyond a few days.}
    \label{fig:loo}
\end{figure}

The GP prediction for two supernovae is shown in Figure~\ref{fig:loo}. 
In this figure, the upper panels show the wavelength-averaged flux for each phase (which is close to the bolometric lightcurve).  The lower left panel shows an example of a successful series of predictions: 
For each SN, we remove each epoch in turn and redo 
the GP model and predictions.  The removed spectrum, in black, is compared to
the GP prediction for its date, in color.  (We call this a ``leave-one-out
test.'')  The fact that the removed spectra can all be faithfully reproduced
indicates that the GP is able to predict this supernova's spectra at any phase
accurately. This was the case for the vast majority ($\sim$83\%) of the supernovae.  The rarer exceptions, as shown in the right panels, are discussed in the following section.

\begin{deluxetable}{lll}
    \tablecaption{Summary of sample selection cuts.
    \label{tab:cuts}
    }
    \tablecolumns{3}
    \tablewidth{0pc}
    \tablehead{Selection requirement &       Included SNe & Included spectra}
    \startdata
    Initial sample & 80 & 833 \\
    Phase $<25$ & 80 & 683 \\
    First phase $<+2$ days & 70 & - \\
    Phase gap $<10$ days & 67 & - \\
    $\ge5$ spectra & 66 & 591 \\
    GP prediction quality & 55 & 468 \\
    \midrule
    Full time series & 55 & 468 \\
    Near-maximum & 49 & 420  \\
    \enddata
    \vspace{-0.5cm}
    
    \tablecomments{All of the selection requirements are applied to individual SNe, except for the ``Phase $<25$'' cut, which applied to individual spectra only, and the ``GP prediction quality'' cut, which is applied to both individual SNe and spectra. Refer to the text for more details on
    individual selection requirements.
    }
\end{deluxetable}

\subsection{Sample Selection}\label{sec:sample}

Demonstration of a reduction in dispersion of the standardized
peak magnitudes of SNe~Ia requires a sufficient
number of SNe~Ia to
measure that dispersion with an uncertainty of $\sim$0.01~mag.
This would require at least $N \sim 50$~SNe~Ia, if the dispersion is $\sigma \sim$ 0.1~ mag, as seen for the best standardization techniques discussed above, and if the accuracy on the dispersion scales as $\sigma/\sqrt{2N}$.
Thus we commenced
this analysis as soon as the available Nearby Supernova Factory sample reached this approximate size. 
Once the sample was selected, it was frozen and no further supernovae were added to the sample, as required for a blind
analysis. Here we present the details of the sample selection.

As the goal of this study is to measure a dispersion that we
anticipate could be much smaller than the canonical $\sim15$\%, it
is important to only include SNe~Ia having sufficient sensitivity.
Broadly speaking, this requires good signal-to-noise, accurate flux
calibration and host galaxy subtraction, and temporal sampling
sufficient to avoid GP prediction errors. Host galaxy peculiar
velocities must be a small fraction of the Hubble expansion velocities
so that redshift is a reliable indicator of relative distances. In 
general, we apply the types of cuts typical of SN cosmology analyses
using lightcurve-based standardization since these are what we will
be comparing to.

The bulk of the SNfactory dataset is in the range $0.03<z<0.08$.
Comparison of two SNe~Ia at $z=0.03$ having peculiar velocities of
300~km/s will result in a uncertainty on their relative brightnesses
of 14\%, so we must exclude the handful of SNe~Ia below this redshift.
We also set an upper limit of $z=0.1$ since the few SNe above this
redshift were observed with much lower S/N than the main program.

Next, we require spectra to have high-quality flux calibration, as
described in \citet{buton13}. This calibration is provided by
observations of multiple spectrophotometric standard stars during
the night, and on non-photometric nights, a measurement of the loss
due to clouds using the SNIFS parallel imaging channel.  At least
two photometric nights per spectral time series were required in
order to ensure adequate reference brightnesses from the parallel
imager.  We also required final reference datacubes taken under
conditions of good seeing, and sufficient for the host galaxy
subtraction method presented in \citet{bongard11} to converge to a
proper solution over the entire portion of sky and host galaxy
sampled by each spectral time series. At the time of the start of
this analysis, 80 of our SNe~Ia satisfied these requirements.

For cosmological analyses where lightcurves are fit using templates or a
model, it is typical to require at least 5 lightcurve points, with
the first starting no later than 6~nights after $B$-band maximum light
\citep[e.g.][]{suzuki12}.  Such lenient requirements are possible
because the templates and models are very stiff and so do not require
that all phases be strongly constrained. Because we do not rely on
templates or models, and must produce GP predicted spectra for
epochs other than those directly observed, our coverage cuts must
be more stringent. We perform initial fits with the SALT2.4 lightcurve fitter \citep{guy07a, betoule14} to BVR
photometry synthesized from the spectrophotometry in order to
determine the phases for our observations, and then require that
the first spectrum be no later than 2.5~nights after maximum. This
requirement results in only very small extrapolation to maximum for
SNe~Ia whose coverage begins after maximum.
We restrict our study to spectra
taken before 25~nights after maximum; after this phase SNfactory switched
to a much less frequent follow-up cadence. The final cut on coverage is that there can be no gap in coverage
greater than 10~nights. Though the nominal SNfactory cadence was
2--3~nights prior to a phase of 25 days, occasionally bad weather, 
proximity to the Moon, or
instrument or telescope operational issues would conspire to create
significant gaps. We require a minimum of 5 epochs
passing these cuts.
After these cuts on coverage, 66 SNe~Ia remain.

The above coverage cuts are sufficient for eliminating cases
where GP prediction would be strongly suspect, however in a blind
analysis we would not be able to further reject SNe traced to bad GP predictions after unblinding, so
we deemed it prudent to inspect the predictions more
closely while still blinded. First, we examined the GP predictions for visual quality.
We required that the wavelength-averaged GP predictions (the upper panels of Figure~\ref{fig:loo}) be smooth as a function of phase.  We also performed a leave-one-out test as described
in Section~\ref{sec:interp}.  For supernovae that
fail the leave-one-out test, we drop the problematic epoch.  If it
is not possible to drop the epoch and pass both the leave-one-out
test and previous requirements, we drop the supernova all together.
Two cases fell below the minimum epoch requirement after this
procedure, demonstrating that the temporal distribution is important when
there is a small
numbers of epochs. Six cases had GP hyperparameters
that were not stable under the leave-one-out test, while three others
were on the edge of acceptability but rejected as a precaution. 
Figure~\ref{fig:loo} shows an example of one supernova that passes these cuts, and 
an example of another one with similar coverage that fails.

We are left with a final sample of 55 SNe after this last 
GP prediction robustness check.
For the near-maximum analysis discussed in
Section~\ref{sec:nearmaxtwinness}, we require a spectrum within 
2.5~nights of maximum in addition to the previous data quality cuts. 
For that analysis, this leaves a final sample of 49 SNe. The median
    signal-to-noise of the final sample is $\sim35$ per
    1000~km~s$^{-1}$ interval for spectra near maximum (calculated as in Section \ref{sec:obs}), and
decreases to a median of $\sim15$ per 1000~km~s$^{-1}$ interval for spectra at 20--25 days after
maximum.

Note that all of these criteria are applied to properties of the
observations, not of the supernovae themselves, so this does not
bias the selection towards any particular subpopulation of SNe. 
Additionally, in order to avoid inadvertent
bias, the SN selection methodology was decided upon and implemented
before the analysis results were unblinded. The number of supernovae
and spectra remaining after each stage of cuts is summarized in
Table~\ref{tab:cuts}. 

\begin{figure}[ht]
    \subfigure{
        \includegraphics[width=0.49\textwidth]{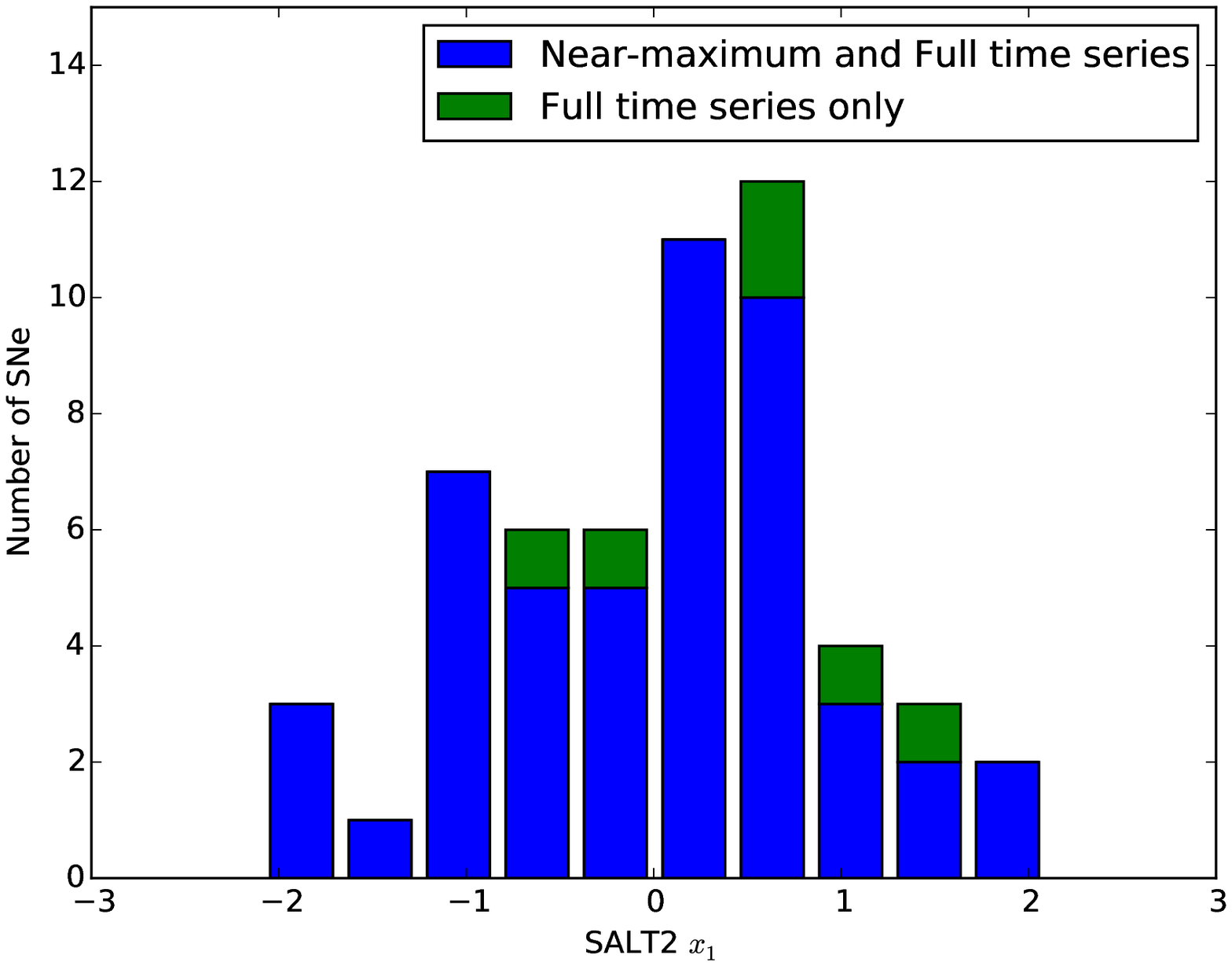}
    }
    \subfigure{
        \includegraphics[width=0.49\textwidth]{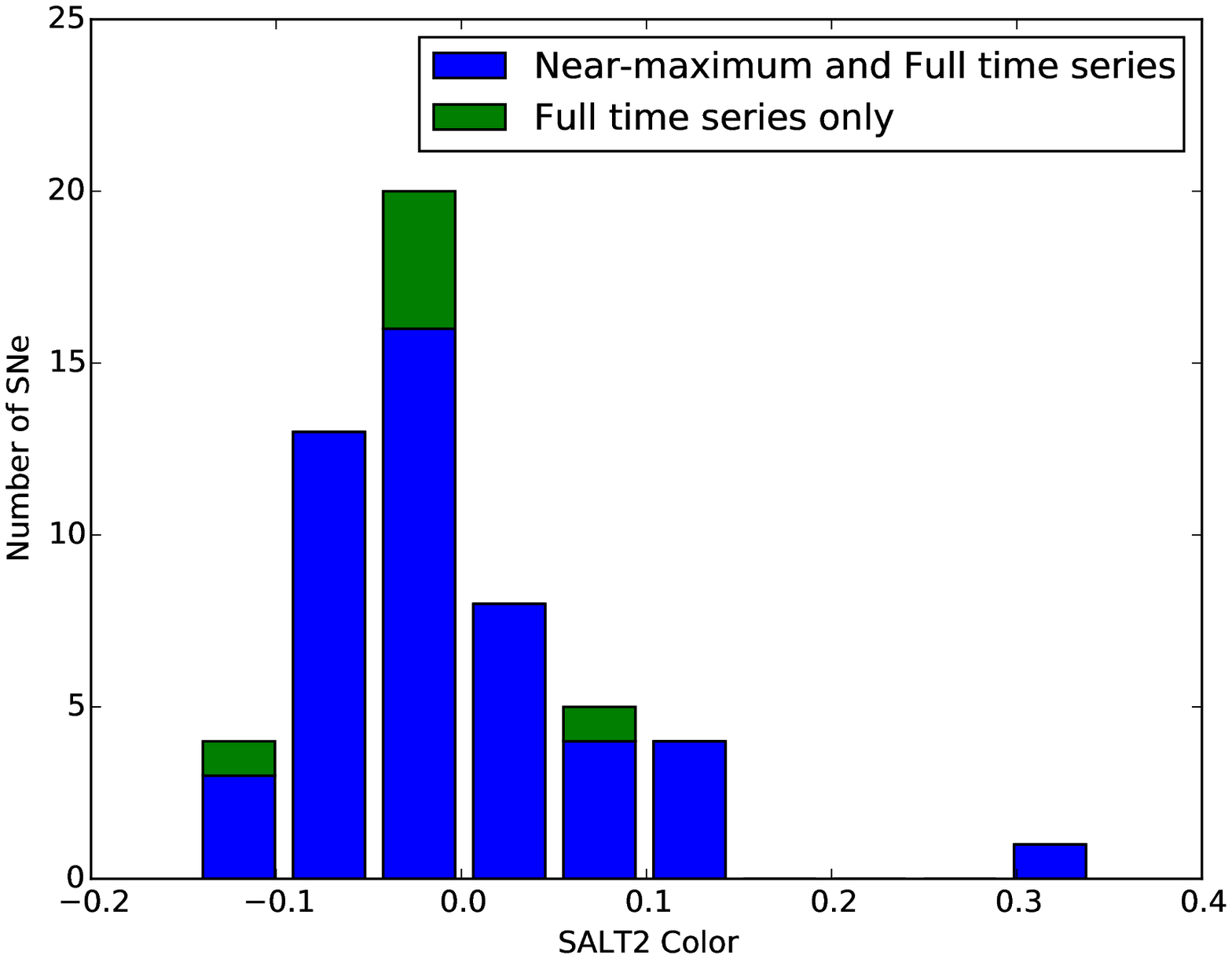}
    }
    \caption{SALT2.4 fit parameters for both the near-maximum dataset 
    and the full time series dataset. Supernovae shown in blue are 
    in both the near-maximum dataset and the full time series dataset.       The supernovae shown in green are those that are dropped 
    from the near-maximum dataset. Left: SALT2.4 $x_1$ distribution.
    Right: SALT2.4 $c$ distribution.}
    \label{fig:salt_dist}
\end{figure}

The resulting sample of supernovae has properties that are 
typical of SNe~Ia used in cosmology analyses. The SALT2.4 
$x_1$ and $c$ distributions for both of these final 
datasets are shown in Figure~\ref{fig:salt_dist}. The 
$x_1$ distribution is roughly Gaussian with the unit 
dispersion expected from the $x_1$ definition. The 
color distribution is a positively-skewed Gaussian, as 
expected --- the result of both intrinsic and foreground 
dust contributions.

\subsection{Preparation for Analysis}

As a further safeguard against inadvertent bias, the dataset has been 
divided into training and validation
subsamples.  The training and validation subsamples were constructed
to have similar distributions in lightcurve parameters and host-galaxy
properties.
We performed all of the initial exploration only on the training
sample.  When the analysis was finalized, it was applied to the
validation sample to act as an independent test and ensure the
result was not simply tuned to the training sample. 
Of the 55 SNe
that pass the cuts, 28 are in the training sample, while 27 are in
the validation sample. For the near-maximum analysis, 24 are in the
training sample, while 25 are in the validation sample. For the full sample,
there are 1309
independent pairs of SNe matching our requirements in the full time series
sample and 1009 pairs in the near-maximum sample.

Prior to analysis all spectra are first corrected for Galactic extinction
based on the \citet{sfd98} dust map, as provided by the NASA Extragalactic Database. Because the SNfactory program specifically avoided SNe with large
Galactic extinction, these corrections are small and the uncertainties associated with this correction are negligible. Next,
in order to compare spectra, we shift the spectra of all SNe to a common
arbitrary redshift in both brightness and wavelength.
The redshifts used for this purpose are taken from \citet{childress13a}. Brightnesses
are brought to a common system assuming a $\Lambda$CDM cosmology with $\Omega_M = 0.28$ and $\Omega_\Lambda = 0.72$. The specific choice of
cosmology is not important since we are working at low redshift where the expansion is linear. Along with this shift to a common redshift, we bin
the spectra into 1000 km/s bins from 3300 \r{A} to 8600 \r{A}. This gives a
total of 288 bins per spectrum (a resolution of R~$\sim150$).

\section{Pairing and Ranking}\label{sec:rank}

With our selection of supernovae completed, we develop a method for pairing the
supernovae and ranking those pairings, from good twins to non-twins.  This
method is developed on the training half of the sample before unblinding the
results, and then applied to the full sample.

\subsection{Near-Maximum Twinness}\label{sec:nearmaxtwinness}

For the first analysis in this paper, we calculate the twinness using 
only the spectrum with phase closest to maximum for each SN.
For each SN in the 
near-maximum sample, we compare this spectrum with the GP prediction 
of every other supernova at that spectrum's phase.
The goal in comparing supernovae to each other is to find pairs that are
spectroscopically similar. Any such pairs would be considered
``twins.'' For our definition of twinness, we choose to
use what is effectively a $\chi^2$ fit with an error floor. We fit 
for extrinsic differences between the two supernovae being compared, and 
call the resulting pseudo-$\chi^2$ the ``twinness''. Mathematically, we define twinness as follows:
\
\begin{equation}
    \xi(p_i) = \frac{1}{N_{DOF}}\sum_{\lambda_j} \frac{[f_A(p_i, \lambda_j) - \alpha(\lambda_j)~f_B(p_i, \lambda_j)]^2}{\sigma^2_A(p_i, \lambda_j)+ \alpha^2(\lambda_j)~\sigma^2_B(p_i, \lambda_j) + (\gamma~f_A(p_i, \lambda_j))^2}
\label{eqn:xi}
\end{equation}

where $f_A$ ($\sigma_A^2$) are the data flux (variance) of SN$_A$ and $f_B$
($\sigma_B^2$) are the GP-predicted flux (variance) of SN$_B$. $N_{DOF}$ is the number of degrees of freedom, which is equal to 286 for this analysis (288 wavelength bins with 2 fit parameters). This equation is
only defined for a single phase $p_i$. $\lambda_j$ represent the different
wavelength bins, and the sum is over all bins.

A definition of twinness should be made independent of
extrinsic dust reddening since it is not a property of the supernova
itself. To take dust extinction
into account we use the parameter $\alpha$, which is defined as follows:
\begin{equation}
\alpha(\kappa,\Delta E(B-V),\lambda_j) =
    \kappa~10^{-0.4(a(\lambda_j) + b(\lambda_j) / R_V)R_V \Delta E(B-V)}
\end{equation}
where the scale factor $\kappa$ captures any remaining overall brightness difference.
$\Delta E(B-V)$ captures a Cardelli-like color difference, with
$a(\lambda)$ and $b(\lambda)$ defined in \citet{cardelli89a}. $R_V$ is fixed
to 3.1 for the initial analysis, and 2.8 is used for a later analysis variant.

For each pair of supernovae, $\kappa$ and $\Delta E(B-V)$ are free
parameters fit by minimizing the twinness, $\xi(p_i)$. Twinness fits
are performed in both directions, with the GP-prediction of SN$_B$ at 
the phase of SN$_A$ and then the GP prediction of SN$_A$ at the phase of 
SN$_B$. We perform the fits separately because the different phase
sampling of the two lightcurves will result in different GP predictions
depending on the direction of the comparison. We find that the two values of
$\xi(p_i)$ rarely differ by more than $\sim 10\%$.
For each SN pair, the two values of $\xi(p_i)$ and the two values of
$\Delta E(B-V)$ are each averaged. Similarly a single value for
$\kappa$ is derived as the geometric mean:
\begin{equation}
    \kappa'_{AB} = \sqrt{\frac{\kappa_{AB}}{\kappa_{BA}}}
\end{equation}

The remaining parameter in Equation~\ref{eqn:xi} is $\gamma$, 
representing an empirically-determined error floor. Without this 
term, the definition 
of $\xi(p_i)$ will favor pairs of lower signal-to-noise SNe 
and disfavor high signal-to-noise SNe having even minor spectral 
differences. $\kappa$ and $\Delta E(B-V)$ are initially fit for each pair 
with $\gamma=0$. We then introduce a finite $\gamma$ in 
the final twinness fit so that different pairs can be compared 
to each other without a strong dependence on their signal-to-noise.
We tested the addition of error floors from 
$\gamma = 0.02$ to 0.20 in steps of 
0.02 on the training sample. We found that $\gamma = 0.12$ produced a ranking
of pairs for which the best 50 pairs had a signal-to-noise distribution most
similar (evaluated using a Kolmogorov-Smirnov test) to that of the full
training set.

\begin{figure}
    \subfigure{
        \includegraphics[width=0.48\textwidth]{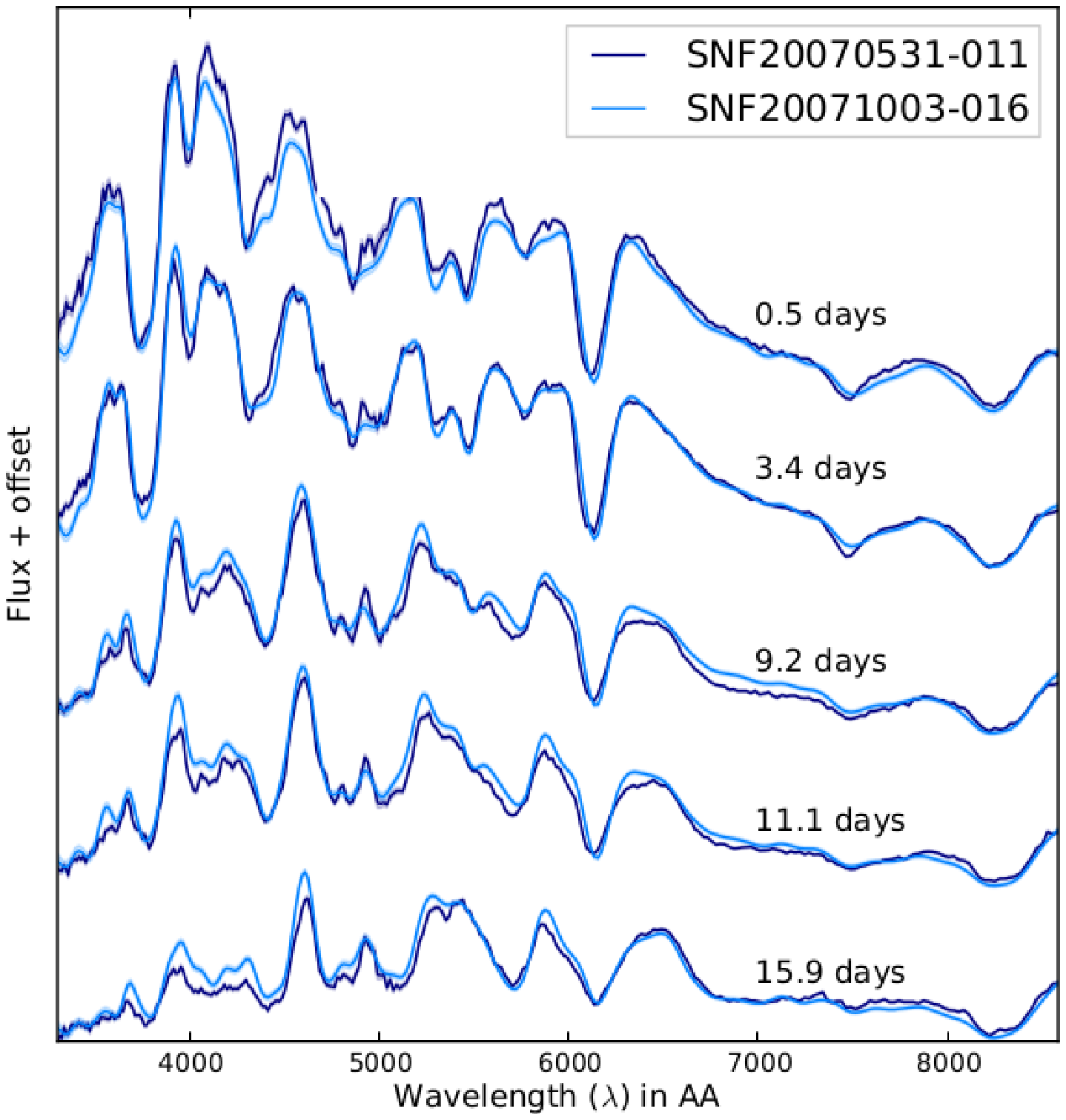}
    }
    \subfigure{
        \includegraphics[width=0.48\textwidth]{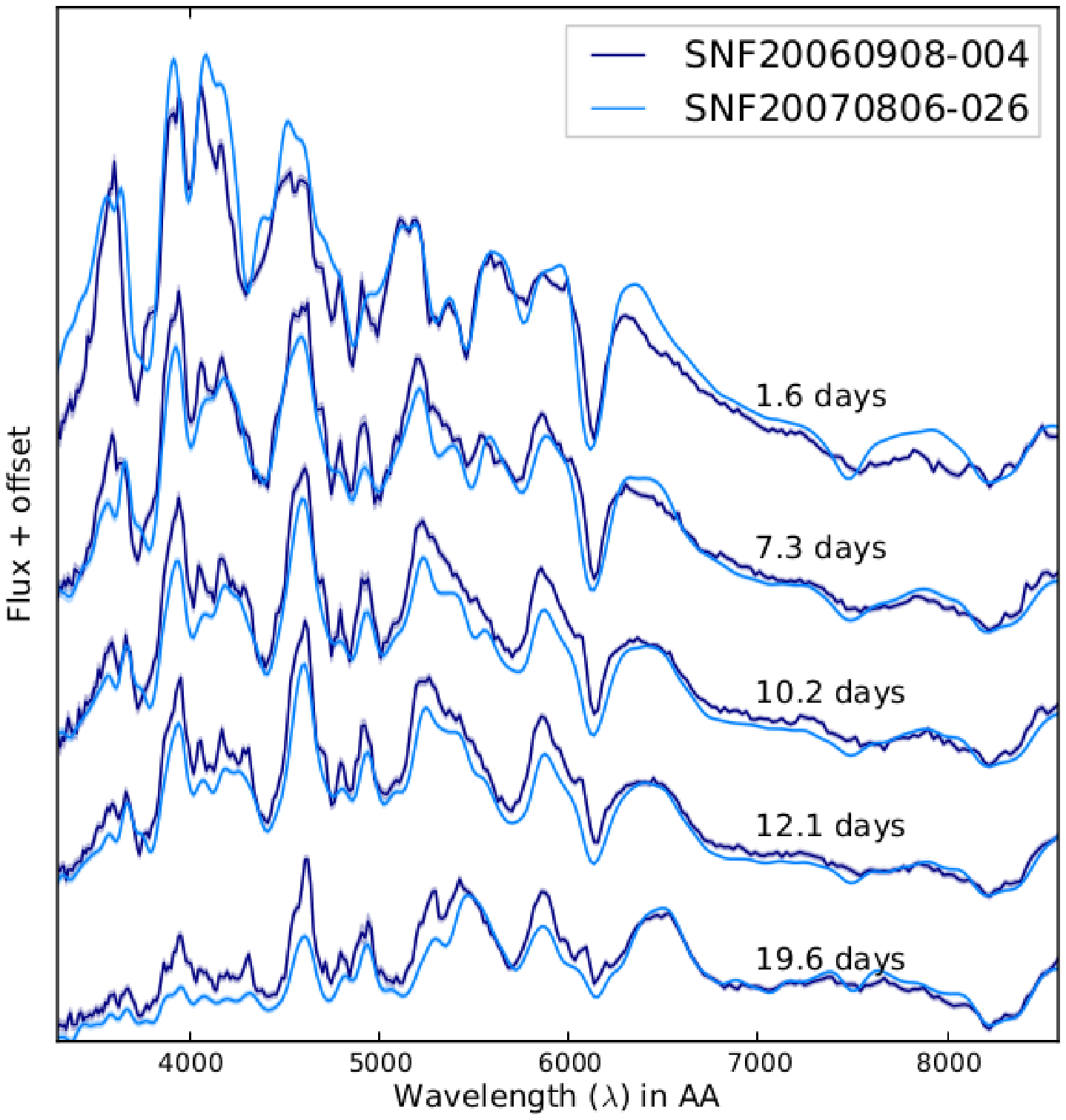}
    }
    \caption{Examples of randomly selected good and bad pairs of supernovae
    from the training set. 
    Data for the first supernova is shown in dark blue, and the GP 
    prediction for the second one at the phases of the first is shown 
    in light blue. Left: SNF20070531-001 paired with SNF20071003-16.
    The near-maximum $\xi$ value for this pair is 0.430, which 
    corresponds to the 6th percentile of pairings, low enough to 
    be considered a twin. Right: SNF20060908-004 paired with SNF20070806-026. 
    The near-maximum $\xi$ value for this pair is 1.755,
    which corresponds to the 77th percentile of pairings 
    --- too high to be considered a twin.
    }
    \label{fig:twin}
\end{figure}

These $\xi$ values give an order to the pairs from ``best pairs'' to
``worst pairs''. Figure~\ref{fig:twin} shows both a randomly chosen pair out of
the best 10\% of ordered pairs and a randomly chosen pair out of the worst 50\% of ordered pairs. 
The level of spectral agreement for the good pair, shown in the left panel,
is impressive, much better than current
theoretical models of supernova explosions are able 
to achieve \citep[e.g.][]{roepke12}. The bad twin example illustrates the 
lack of spectral agreement in
the non-twins regime. Note that, while we show the data at all phases, for
the near-maximum analysis the
twinness value was calculated using only the spectra closest to
maximum.

\
Since our data are spectrophotometric, $\kappa$ represents 
the ratio of the absolute brightnesses of the two supernovae. 
$\kappa$ corresponds to a magnitude difference of: 
\begin{equation}
    \Delta M = -2.5\, {\rm log}(\kappa).
\end{equation}
If $\kappa$ is 1, the supernovae have a magnitude difference of zero. 
 The $\xi$, $\Delta M$ and $\Delta E(B-V)$ values for each pairing of
these supernovae are shown in Appendix~\ref{sec:full_data}. The
$\Delta E(B-V)$ values are consistent with differences between SALT2 colors,
shown in Figure \ref{fig:salt_dist}, with a Pearson correlation coefficient of
0.90.

\subsection{Full Spectral Time Series Twinness}
\label{sec:fullts}

As a variant of this analysis, we also define and
calculate values for a twinness metric that operates 
on the full spectral time series, rather than just the 
spectrum closest to maximum light. Unfortunately,
implementation errors were found in
this full time series analysis after unblinding that had 
to be fixed  after unblinding. Although we ran this method 
with only those specific errors fixed, we nevertheless do 
not consider our results on the full spectral time series
to be blinded. Since the implementation errors are unrelated
to any properties of SNe, this portion of the analysis
is unlikely to be biased despite not being blind.

In order to make the most of the information from different
phases, it is necessary to determine the best relative weights for each phase.
The full covariance matrix between all measurements in both phase and wavelength would be ideal for this purpose, but we do not have a way to measure this covariance matrix directly. As an approximation, we estimate $\Delta M$ at each phase using the method described for the near-maximum sample, and we calculate the covariance matrix $C(p_i, p_k)$ between these estimates of $\Delta M$. We then take a mean of the pseudo-$\chi^2$ values from Equation~\ref{eqn:xi} weighted by this covariance matrix to generate an overall pseudo-$\chi^2$ function, as follows:
\
\begin{equation}
    \label{eqn:upweighting}
    \xi = \frac{1}{N_{DOF}}\frac{\sum_{i,k} C^{-1}(p_i, p_k) \xi(p_k)}{\sum_{i,k} C^{-1}(p_i, p_k),}
\end{equation}
which we then minimize.
Here $\xi(p_k)$ is the same as in Equation~\ref{eqn:xi},
but without the $N_{DOF}$ term (the $N_{DOF}$ term in Equation \ref{eqn:upweighting} is now $N \times 288 - 2$ where $N$ is the number of phases used, 288 is the number of wavelength bins and 2 is the number of fit parameters).
$C(p_i, p_k)$ represents the covariance of the dispersion in brightness between phases $i$
and $k$. Application of this equation relies on the ability to calculate the matrix, $C(p_i, p_k)$,
which is actually a two dimensional function since all of the
different supernovae were measured at different phases.
Thus, it is necessary to evaluate this function for arbitrary $p_i$ and $p_k$. The details of estimating this
matrix are somewhat technical, and can be found in Appendix~\ref{sec:weightcov}. From the methodology given
there, $C(p_i, p_k)$ is calculated and used to execute fits to minimize
$\xi$ in each direction and thereby obtain a single value of $\kappa$ and $\Delta E(B-V)$ per pairing, as before.

\section{Brightness Difference Methodology}\label{sec:deltam}

Having completed the ordering of supernova pairs, we can assess
the extent to which more twin-like pairs (those with low $\xi$) 
have more similar brightnesses. In this section we describe the 
measurement of relative brightnesses and the results for the 
near-maximum and full spectral time series cases. We then describe 
consistency checks that were performed to validate and better 
understand the results.

\subsection{Magnitude Calculation and Error}

Recall that when comparing one 
supernova to another we fit for a scale factor, $\kappa$, which we convert into a brightness difference $\Delta M$.
The uncertainty on this brightness difference is determined from the uncertainty on
the fit of $\kappa$, and external uncertainties due to both the host-galaxy peculiar velocities (we assume a 300 km/s uncertainty), and the per-spectrum random gray calibration inferred from standard star measurements \citep{buton13}:
\
\begin{align}
    \sigma_{\Delta M}^2 &= \sigma_{fit}^2 + \sigma_{ext}^2 \\
    &=  \sigma_{fit}^2 + \sigma_{\Delta v_A}^2 +
    \sigma_{\Delta v_B}^2+ \sigma_{calib}^2
    \label{eqn:dmerr1}
\end{align}

where
\begin{equation}
    \begin{split}
        \sigma_{fit}& = (2.5/\ln(10))\, (\sigma_\kappa/\kappa) \\
        \sigma_{\Delta v}& = (5 / \ln(10)) \times (300/cz) = 0.00217 / z\\
        \sigma_{calib}& = 0.025
    \end{split}
    \label{eqn:dmerr2}
\end{equation}
Note that even with the low redshift cut-off we have imposed, the fit uncertainties ($\sigma_\kappa$ and $\sigma_{\Delta E(B-V)}$)
are subdominant to the peculiar velocity uncertainties, which comprise roughly
80\% of the measurement error budget.

\begin{figure}[ht]
    \centering
    \includegraphics[width=0.75\textwidth]{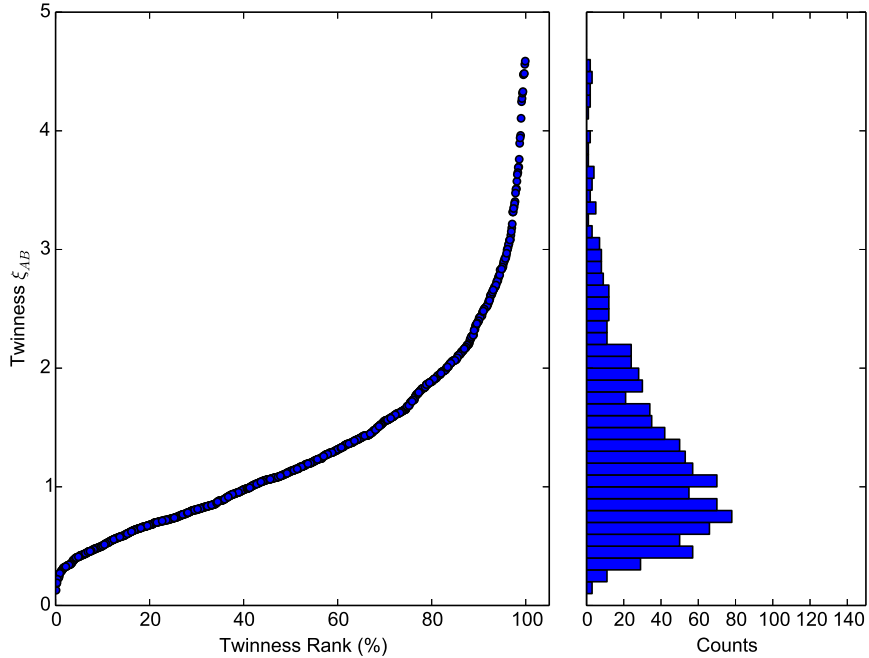}
    \caption{\label{fig:f3022} Near-maximum $\xi$ distribution for $R_V=3.1$.
    Due to the long
    tail of the $\xi$ distribution, we work with the percentile in $\xi$ instead
    of the raw value itself. Left: Conversion between $\xi$ and $\xi$ percentile.
Right: histogram of the near-maximum $\xi$ twinness values. }
\end{figure}

\subsection{Near-maximum results}\label{sec:nearmaxres}

With these considerations
in mind, the distribution of near-maximum twinness values 
$\xi$ for $R_V=3.1$
is shown in Figure~\ref{fig:f3022}. The histogram on the 
right shows that there
is a broad peak in
twinness with a long tail to high values. In order to 
include this
tail, we use the twinness percentile for further analysis 
instead of the raw
twinness score. The pairs of supernovae are then divided 
into 17 bins of equal
twinness percentile, each with 59-60 pairs (17 was 
arbitrarily chosen because
the full spectral time series sample has 1309 pairs, and 17 divides 1309 evenly;
it has no deeper meaning and was not fine-tuned in any way).
Within each bin the $\Delta M$ values are given uniform weight
when calculating the bin dispersions and their uncertainty.

\begin{figure}
    \centering
    \includegraphics[width=0.49\textwidth]{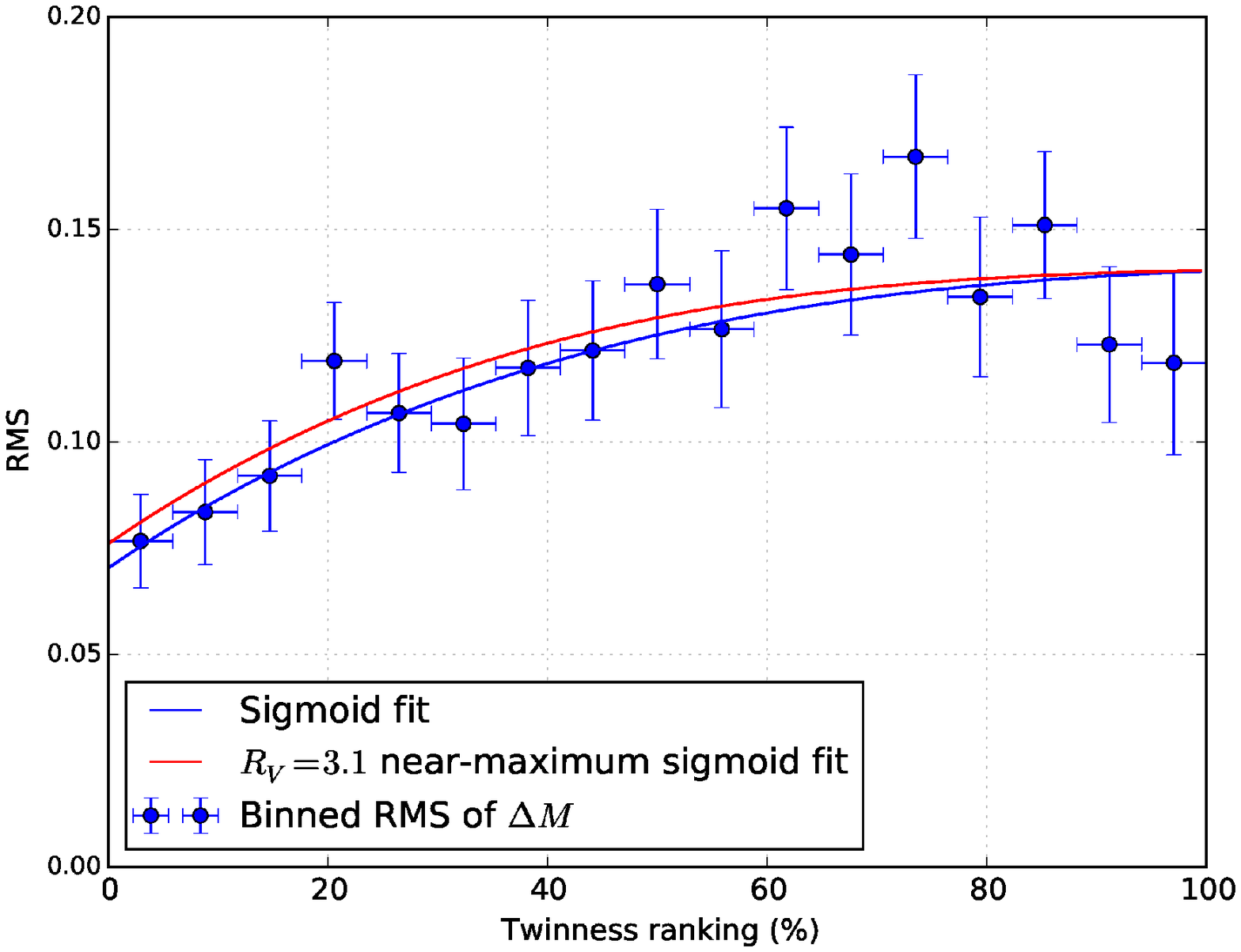}
    \includegraphics[width=0.49\textwidth]{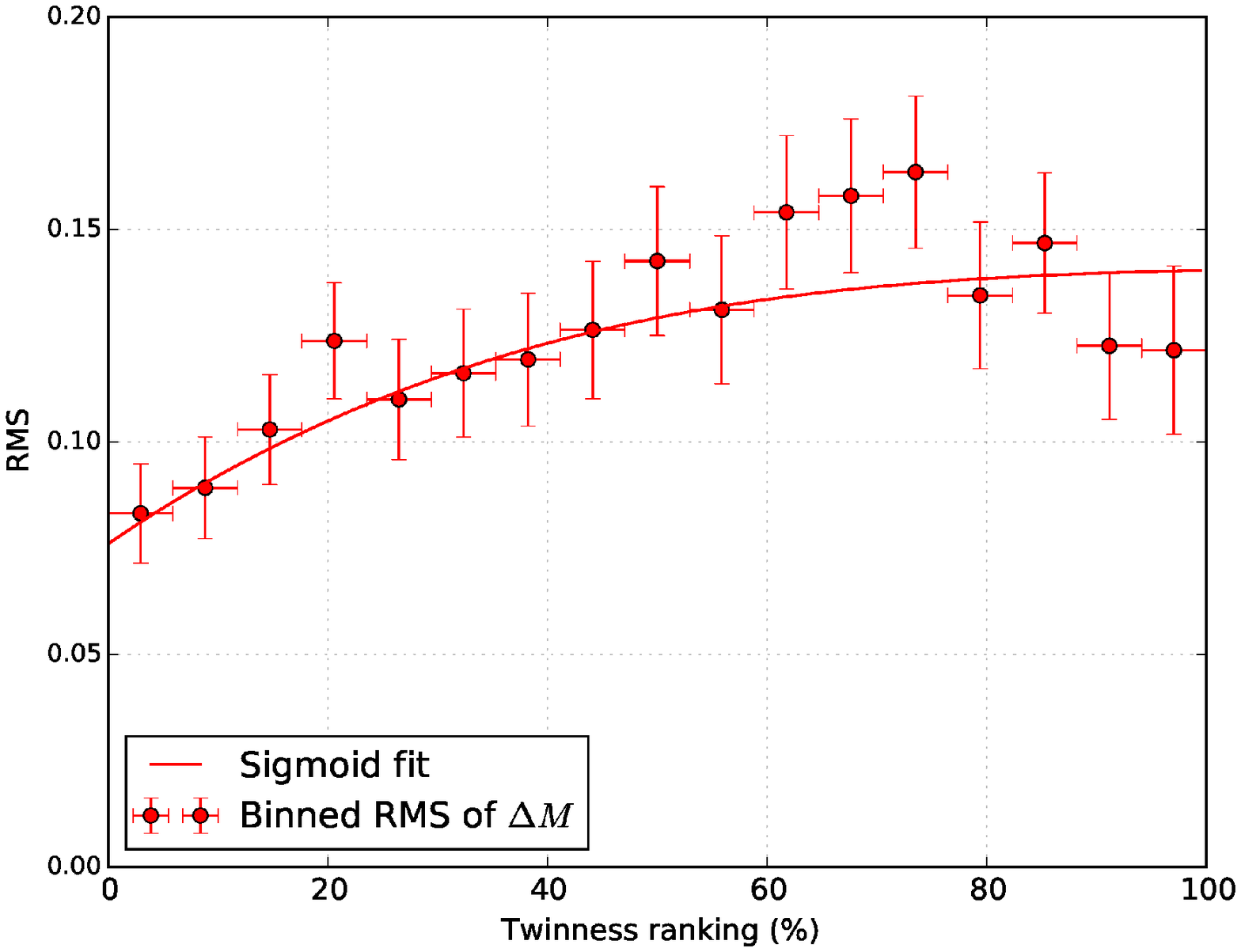}
    \caption{\label{fig:f3050} Binned unweighted RMS vs. twinness
    percentile for the near-maximum analysis. Left: $R_V=2.8$, Right: $R_V=3.1$. The
    data is binned by twinness rank, and the data points shown are the unweighted RMS
    values in each bin. On the left plot, the fit from the main $R_V=3.1$
    result is shown as a red line for comparison.}
\end{figure}

The main result of the twins analysis, using the combined training
and validation sample, is shown in Figure~\ref{fig:f3050}. 
On these plots we give the ensemble SN magnitude RMS, which is the appropriate metric for direct comparison with other SN standardization approaches, in each of the 17 pre-defined twinness percentile bins. The measurement errors are small and sufficiently uniform that an unweighted RMS over all values of $\Delta M/\sqrt{2}$ in each bin accurately accounts for the contribution of each SN to the ensemble RMS in that bin. We ran the blind 
analysis using an $R_V$ value of 3.1, however we have reason to believe
that an $R_V$ of 2.8 \citep{chotard11} may be more appropriate. We report 
all numbers for both of these values of $R_V$, but we consider the 
$R_V=3.1$ analysis to be our pure blinded result. There are few 
differences between the results with the two $R_V$ values, so the following analysis
applies to both.

The bin occupancies for each supernova are shown in Appendix~\ref{sec:bin_occupancies}.
There is a strong correlation between the dispersions of nearby bins due to the
fact that any particular supernova can be included in multiple bins. Errors are
estimated on the unweighted RMS in each bin using a Monte Carlo simulation to generate the
covariance matrix. We discuss this simulation in Appendix~\ref{sec:monte_carlo}. We fit a sigmoid model to the unweighted RMS values in each bin with
the following form:
\begin{equation}
    f(x) = \sqrt{y_0^2 + ( (y_{100} - {y_0} )^2 - y_0^2) \frac{\textrm{erf} (x
    k)}{\textrm{erf}(100 k)}}
\end{equation}

The parameters in this expression characterize the observed rise
in the RMS vs. $\xi$ data. $y_0$ represents the 
dispersion for the best twins, 
$y_{100}-y_0$ represents the difference in dispersion between the
worst and best twins and $k$ affects the shape of the rise. We 
calculated the significance of observing a decreased dispersion for
better twins by
comparing the $\chi^2$ statistic for both the sigmoid fit and a constant 
fit using
the Bayesian Information Criterion \citep{schwarz78}. The Bayesian Information 
criterion is calculated as BIC $= \chi^2 + \nu \log (n)$ where $n$
is the number of data points (17) and $\nu$ is the number of free 
parameters in the fit (3 for the sigmoid, and 1 for the flat line). 
A $\Delta$BIC larger than 10 corresponds to very strong evidence 
that one model is favored over another. The results of this comparison
are presented in Table~\ref{tab:significance}. For both $R_V$
values, the BIC values imply that the improvement in the unweighted 
RMS for better twins is statistically significant.

\begin{table}
    \caption{Change in Bayesian Information Criterion of
    near-maximum twin
    result for constant and sigmoid fits}
    \vspace{0.5em}
    \centering
    \label{tab:significance}
    \begin{tabular}{lcc}
    \toprule
    {} &       $R_V=2.8$ &       $R_V=3.1$ \\
    \midrule
    $\chi^2$ for constant & 37.90 & 33.69 \\
    $\chi^2$ for sigmoid & 16.17 & 13.51 \\
    $\Delta$BIC & 16.06 & 14.51 \\
    \bottomrule
    \end{tabular}
\end{table}

We now focus on the dispersion in the best twinness bin. In principle, with a large reference sample, 
any new supernova will have many close twins in the
reference sample and hence many pairs in this best twinness bin, and
these would be the only other supernovae that would be used to determine its
relative distance.
The quoted values for the unweighted RMS thus correspond to 
this limiting case.
The values and their uncertainties calculated for these 
dispersions are presented in Table~\ref{tab:dispersion}. 

\begin{deluxetable}{lllllll}
\tabletypesize{\footnotesize}
\tablecolumns{7}
\tablewidth{0pc}
\tablecaption{Brightness dispersion for different analyses}
\tablehead{
\multicolumn{1}{l}{Method Variant} &
Percentile &
\multicolumn{2}{c}{With peculiar velocity} & &
\multicolumn{2}{c}{Peculiar velocity removed} \\

& Cutoff & \multicolumn{1}{c}{$R_V$=2.8} & \multicolumn{1}{c}{$R_V$=3.1} & &
\multicolumn{1}{c}{$R_V=2.8$} & \multicolumn{1}{c}{$R_V$=3.1} }
\startdata
Pair-wise Analysis & & & & & \\
\hspace{0.5cm} Near-maximum & & & & & \\
\hspace{1.0cm} {\it Training subset}   & {\it 6\%} &$\mathit{0.073\pm0.020}$ &$\mathit{0.084\pm0.018}$   &  &$\mathit{0.061\pm0.016}$ &$\mathit{0.074\pm0.016}$ \\
\hspace{1.0cm} {\it Validation subset} & {\it 6\%}
&$\mathit{0.088\pm0.019}$ &$\mathit{0.097\pm0.020}\tablenotemark{*}$ &
&$\mathit{0.076\pm0.016}$ &$\mathit{0.085\pm0.018}\tablenotemark{*}$ \\
\vspace{0.1cm}
\hspace{1.0cm} Combined set                                   & 6\%                         & $0.077\pm0.011$                           & $0.083\pm0.012$                             &  & $0.064\pm0.009$                           & $0.072\pm0.010$ \\
\vspace{0.15cm}
\hspace{0.5cm} Weighted full lightcurve                       & 6\%                         & $0.077\pm0.012$                           & $0.085\pm0.011$                             &  & $0.062\pm0.010$                           & $0.072\pm0.010$ \\
Realistic, single-SN distance                                 &                             &                                           &                                             &  & \\
\hspace{0.5cm} RMS$_{nw}$ (as measured)                       & 20\%                        & $0.081\pm0.008$                           & $0.091\pm0.009$                             &  & $0.069\pm0.010$                           & $0.080\pm0.010$ \\
\hspace{0.5cm}  Large reference sample RMS$_{nw}$             & 20\%                        & $0.071\pm0.007$                           & $0.079\pm0.008$                             &  & $0.057\pm0.010$                           & $0.067\pm0.010$ \\
\enddata
\tablecomments{Except where indicated, all results are for the
combined training and validation set of SNe.
}
\tablenotetext{*}{The results of the blinded near-maximum $R_V=3.1$
analyses of the validation subset are indicated with an asterisk.}
    \label{tab:dispersion}
\end{deluxetable}

These results are very promising. The near-maximum study, with
    $R_V=3.1$ (the fully blinded value), yields a dispersion of $0.083\pm0.011$~mag
in the best twinness bin, for the combined
training and validation sample. 38 of the 49 SNe~Ia in the sample are represented in this bin. This is an impressive improvement over the
$\saltrms$~mag dispersion obtained with SALT for exactly the same 
data. In Section~\ref{sec:compare} we show that this dispersion is
among the very best yet achieved among a wide variety of SNe~Ia
standardization methods. It is interesting to note that the results using $R_V=2.8$ (the preferred
value from \citet{chotard11}), appear to be slightly better:
we obtain $0.077\pm0.011$~mag for the dispersion in the best twinness bin.

Since we are working at low redshift, peculiar velocities account for a
significant portion of the spread in magnitude. Therefore it is instructive to
estimate the dispersion that would result without the peculiar velocity
component, as might be seen with a high-redshift SN sample. Using a dispersion
estimate of $\Delta v=300$~km/s for the peculiar velocities and assuming that
they are all independent, we calculate their contribution to the unweighted RMS
and removed it. We find that without peculiar velocities the near-maximum twins
dispersion drops to $0.072\pm0.010$~mag. The results with peculiar velocity
dispersion removed are shown in the right two columns of
Table~\ref{tab:dispersion}.

If we break the supernova sample into its training and validation subsamples,
we find consistent results. For these two subsamples the RMS values in the
first bin are tablulated in the first and second rows of
Table~\ref{tab:dispersion}.  For the training sample the RMS values in the
first bin range from $0.061$ to $0.084$~mag for the different analyses. For the
validation sample this range is $0.076$ to $0.097$~mag. The validation RMS
values are between $0.011$ to $0.015$~mag higher than the training RMS values,
but this is well within the $\sim 0.025$~mag uncertainty between these two
independent subsets of SNe, and thus not statistically significant. Note that
the RMS values of \textit{both} the training and validation subsets are
slightly higher than for the combined sample since SNe are less likely to find
a good twin in a smaller sample (in Section \ref{sec:refsize} we discuss the
further improvement in RMS expected for yet larger samples). The following
results use the combined training and validation sets.

\subsection{Full Spectral Time Series Analysis}

\begin{figure}[ht]
    \centering
    \includegraphics[width=0.49\textwidth]{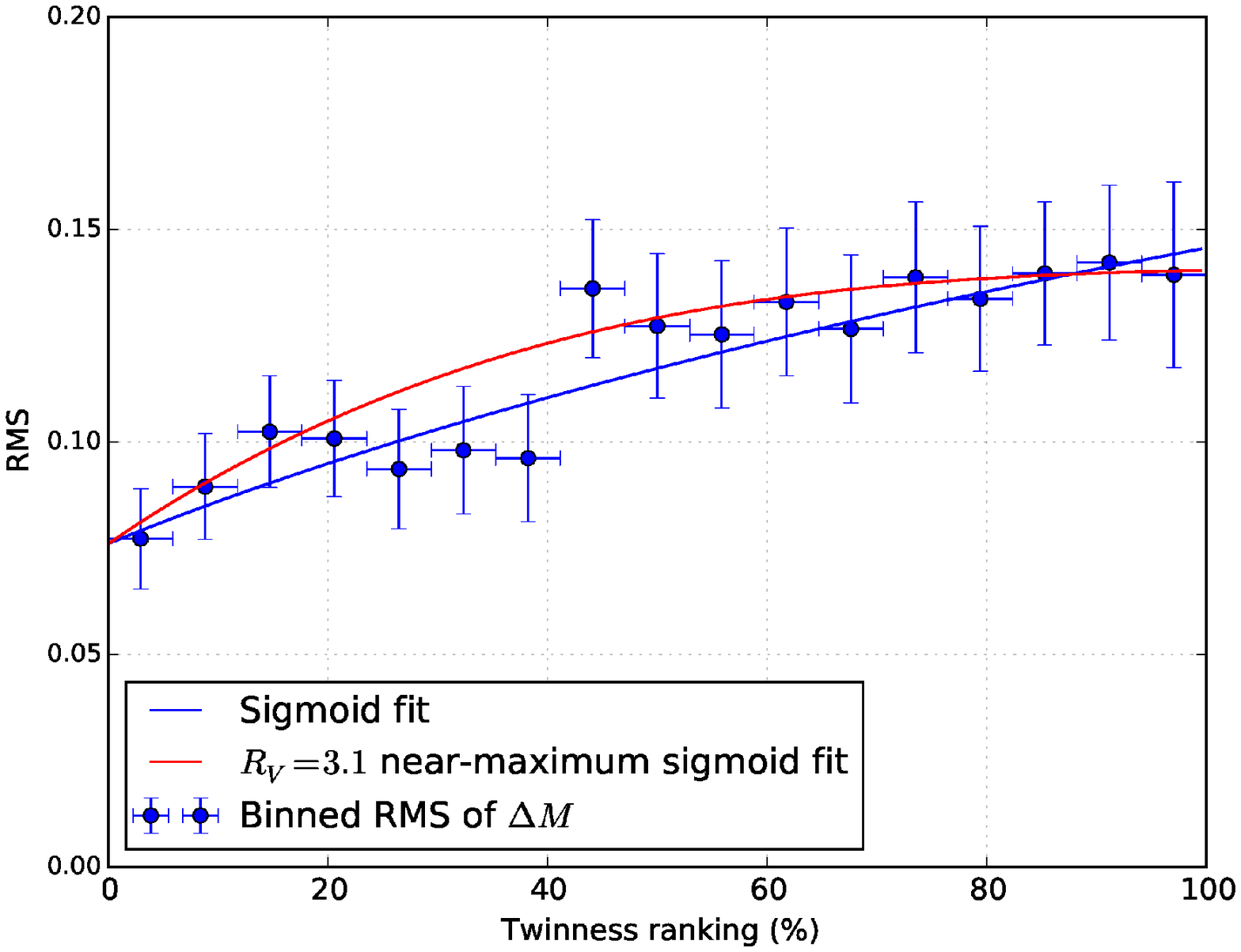}
    \includegraphics[width=0.49\textwidth]{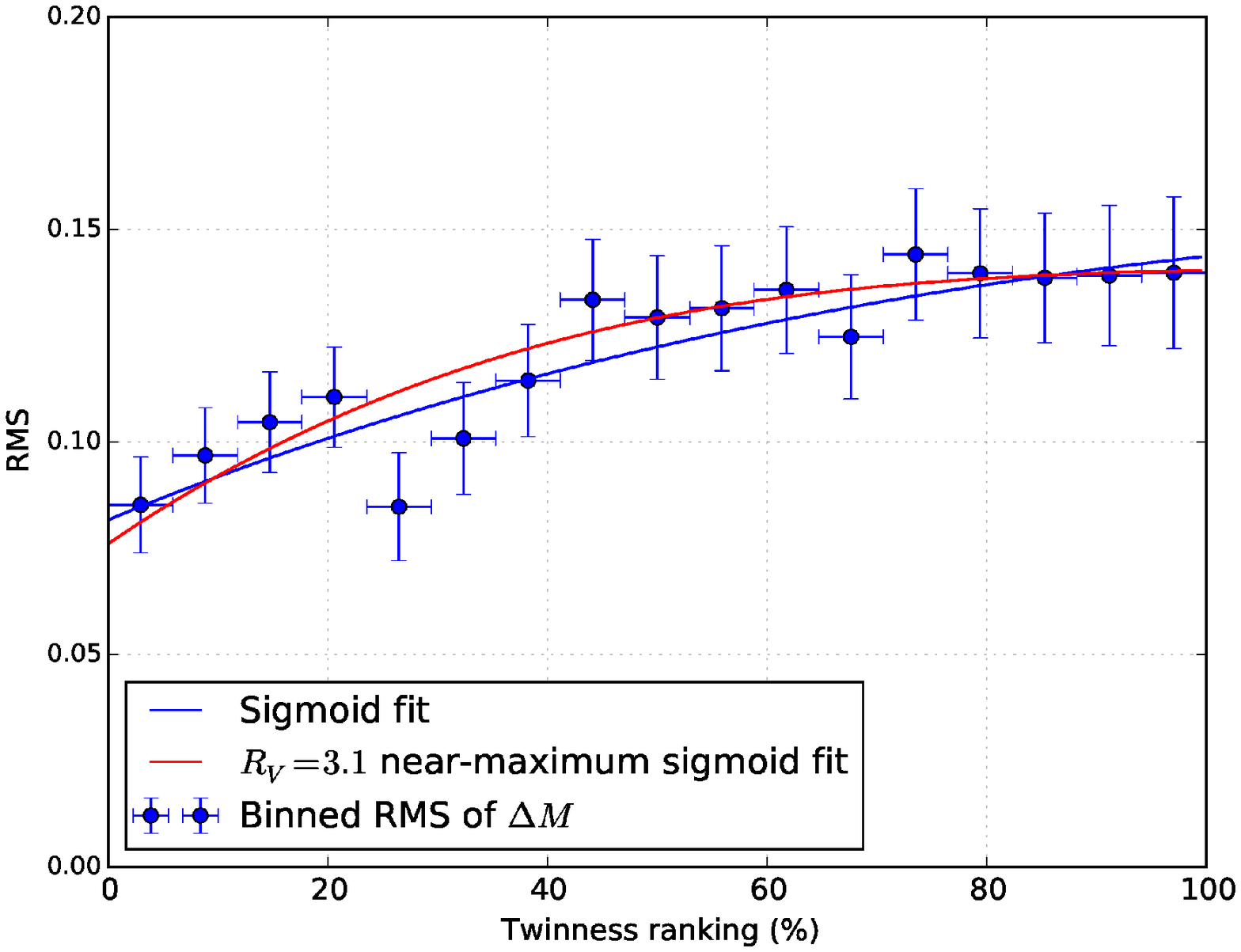}
    \caption{\label{fig:f3071} Binned unweighted RMS vs. twinness
    percentile for the weighted full spectral time series analysis. Left:
    $R_V=2.8$, Right: $R_V=3.1$. The fit from the main
    $R_V=3.1$ near-maximum result is shown as a red line for comparison.}
\end{figure}

The full spectral time series twinness definition from Section~\ref{sec:fullts} 
was used to examine dispersion as a function of twinness, just as for the near-maximum analysis described above. Plots of RMS as a function of $\xi$ for the full spectral time series are shown in Figure~\ref{fig:f3071}.

The full spectral time series twinness values do produce better results
overall than the near-maximum only values. However, most of the gain 
comes from the 20--40th percentile range. The unweighted RMS values for 
the first bins are shown in Table~\ref{tab:dispersion} (43 out of 55 SNe are represented in the
best bin for the $R_V=3.1$ analysis). These RMS values 
agree almost exactly with the near-maximum result indicating that we 
do not gain much, if anything, by including later
phases for good twins. Again, the \citet{chotard11} value of $R_V=2.8$ gives
slightly better results.

\subsection{Consistency}

\begin{figure}[ht]
\centering
\includegraphics[width=0.49\textwidth]{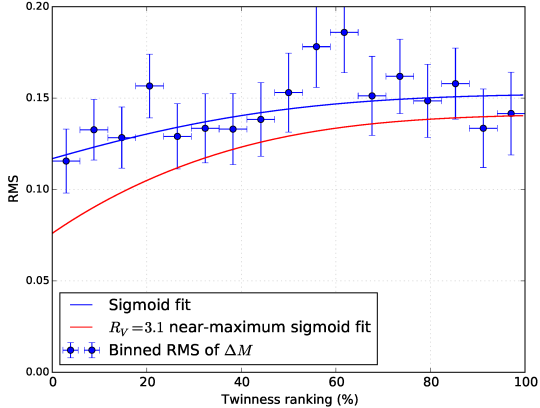}
\caption{\label{fig:f3001} Binned SALT unweighted RMS vs.
near-maximum twinness rank (percentile). The near-maximum 
twins RMS fit with $R_V=3.1$ is shown as a red line 
for comparison. Magnitude differences calculated
with SALT show only a weak dependence on twinness.}
\end{figure}

In this section we present checks on the internal consistency 
of our measurements, and their relation to SALT. First, we examined 
what a traditional standardization would give for the same dataset: we 
calculated corrected magnitudes for each SN using the \citet{tripp98} 
lightcurve width and color relation with $x_1$ and $c$ from SALT2.4 fits. 
The dispersion of these corrected magnitudes is \saltrms.
We investigated whether the twinness can be used to improve the SALT-based
standardized magnitudes, and we show the RMS values using these magnitudes
as a function of twinness in Figure~\ref{fig:f3001}. This figure shows a weak
dependence on twinness, but the results are nowhere near as good as the 
twins method in the best twins bins. This implies that the twins 
analysis is capturing more information
than a conventional standardization based on color and stretch parameters.

Interestingly, the dispersion taken over all bins, i.e. ignoring 
twin rank, is $0.128 \pm 0.012$~mag.
This outperforms the dispersion resulting from the SALT-based standardization.
We believe that this is due to the fact that SALT lightcurve fits 
do not use the full wavelength range and resolution of our spectra. 
To explore this, we created a basic SALT fitter that uses the 
SALT templates over the full spectrum with the full wavelength 
range and same wavelength resolution used in the twins analysis.
Standardization using peak magnitude, $x_1$ and $c$ from this 
fitter gives an overall dispersion of $0.133 \pm 0.013$~mag which is consistent with 
the overall twins RMS. There is no significant dependence of the RMS values
calculated using this method on twinness, so the pairwise method still has a much
better dispersion if only the best twins are considered.
Thus, there appears to be extra information available in a direct comparison 
of two spectra that is not captured in the two parameter model in a SALT-based 
standardization.

We also checked whether the twins result improves when 
removing the remaining correlation of RMS with 
SALT $x_1$ and $c$. This does not improve the result for good twins, 
which is expected since good twins should have very similar 
lightcurve widths, while the color has been taken out as part of
the twinning calculation. It does improve the results somewhat for the
worst twins, where we do not expect the twins method to work well;
the twins method is designed to simply reject those pairings
rather than attempt to match them.

In all the previous analyses, we only calculate the difference in
brightness and colors between two supernovae, never the 
actual values for individual supernovae. As a consistency 
check, we fit for a global set of self-consistent magnitude 
and color differences for each supernova. This gives a 
consistent set of values that are commutative. When comparing 
the $\Delta M$ and $\Delta E(B-V)$ values from the global 
fits to those calculated with the twins method, the 
difference $\Delta M_{\text{fit}} - \Delta M_{\text{twins}}$
has an RMS of 0.030~mag and the difference $\Delta E(B-V)_{\text{fit}} - \Delta
E(B-V)_{\text{twins}}$ has an RMS of 0.0095~mag. These values are
small compared to the sample dispersions and indicate good agreement.

\section{Discussion}\label{sec:discussion}

We now turn to several considerations concerning the application 
of the twins method for cosmology. We begin by comparing our 
result with the dispersions and efficiencies attained by other
SN~Ia standardization techniques. Next, we explore how realistic
distance determinations can be made using the 
twins method, and whether we can obtain distances with uncertainties
as small as the dispersions discussed in Section \ref{sec:deltam}. 
Finally, we consider how the uncertainty on these distance determinations
is expected to scale with the size of the reference sample.

\subsection{Comparison to Other Standardization Techniques}
\label{sec:compare}

For the near-maximum twins analysis we found a dispersion of 
$0.083\pm0.012$~mag using $N=38$ of the 49 SNe~Ia in the sample.
This is a substantial improvement over the
SALT dispersion of $\saltrms$~mag found using the same data.
The literature contains a number of results showing small
dispersion, to which our result can be compared.
For this purpose, we quote the best dispersion presented in 
each paper. To be consistent with our values, no measurement 
uncertainties were subtracted for these studies. When uncertainty 
on the dispersion was not provided, we calculated
it assuming a Gaussian distribution for the magnitude residuals.

The very best alternative standardization techniques include
the results of \citet{barone12}, who achieved $0.085\pm0.016$~mag 
using NIR lightcurves for all $N=12$ SNe~Ia in their sample;
\citet{lwang03}, who achieved a dispersion of $0.080\pm0.013$ 
using the CMAGIC technique and cuts that included $N=20$ of 
their 40 SNe~Ia; and \citet{kelly15}, who  achieved a dispersion 
of $0.065\pm0.016$ by using SALT2.4 lightcurve parameter standardization
and then selecting the $N=10$ of their 50 SNe~Ia 
having the highest local star formation rate. Our result is 
equally good, but includes many more SNe~Ia than these studies.

Several more alternative standardization techniques report
dispersion values that are higher than, but statistically consistent 
with, ours: \citet{fk11} found a dispersion of $0.109\pm0.010$~mag 
using the $N=65$ of their 121 SNe~Ia having a normal velocity for the
Si~II~$\lambda 6355$~\AA\ absorption line;
\citet{rigault13} obtained a dispersion of $0.105\pm0.012$~mag 
using SALT2.4 and $N=41$ of their 82 SNe~Ia by taking the half of their 
sample having the highest local star formation rate;
\citet{mandel11} obtained a dispersion of $0.101\pm0.019$~mag
using a hierarchical Bayesian model applied to 37 SNe~Ia having 
both optical and NIR imaging lightcurves and $0.113\pm0.016$~mag
using 57 SNe~Ia having both optical and NIR imaging lightcurves.

Thus, the twins method produces a dispersion that is as good, if not 
better than, the best known standardization methods to date, especially considering that these comparison studies were performed unblinded. 
In addition, the efficiency in using SNe~Ia is among the highest 
(78\%), and the size of the sample from which the
dispersion is measured is among the largest.

\subsection{Realistic Application to a New Single Supernova}
\label{sec:highz}

If a set of spectrophotometrically observed high-redshift
supernovae became available, the twins methodology could be
applied (the currently considered design for the WFIRST space telescope mission
\citep{spergel15} could provide such a sample). A large sample of
low-redshift SNe~Ia adjusted to a common
redshift would serve as a set of templates against which the 
high-redshift sample is fit. The brightness 
for each high-redshift SN relative to all the low-redshift 
supernovae it matches well could be calculated using the method
described in Equation~\ref{eqn:xi}. Taking a number-weighted 
average of these $\Delta M$ values would produce an average 
brightness difference, which we denote as $\phi_i$, for each 
high-redshift supernova indexed by $i$. Then $\phi_i$ is a measure of
the relative distance modulus, given by:
\begin{equation}
    \phi_i(\xi_{\rm{cutoff}}) = \frac{\sum_{j, ~\xi_{ij} < \xi_{\rm{cutoff}}}
    \Delta M_{ij}}{\sum_{j, ~\xi_{ij} < \xi_{\rm{cutoff}}}1}
\end{equation}
To mimic this situation within our low-redshift sample, we treat 
each supernova
in turn as though it were the high-redshift supernova. Note that 
because the redshifts have been adjusted to a common redshift in this 
sample, $\phi_i$ would be zero for a perfect standard candle with no 
measurement errors. Hence the spread of those $\phi_i$ values over 
this sample measures the dispersion in this standardization method.

\begin{figure}
    \subfigure{
        \includegraphics[width=0.49\textwidth]{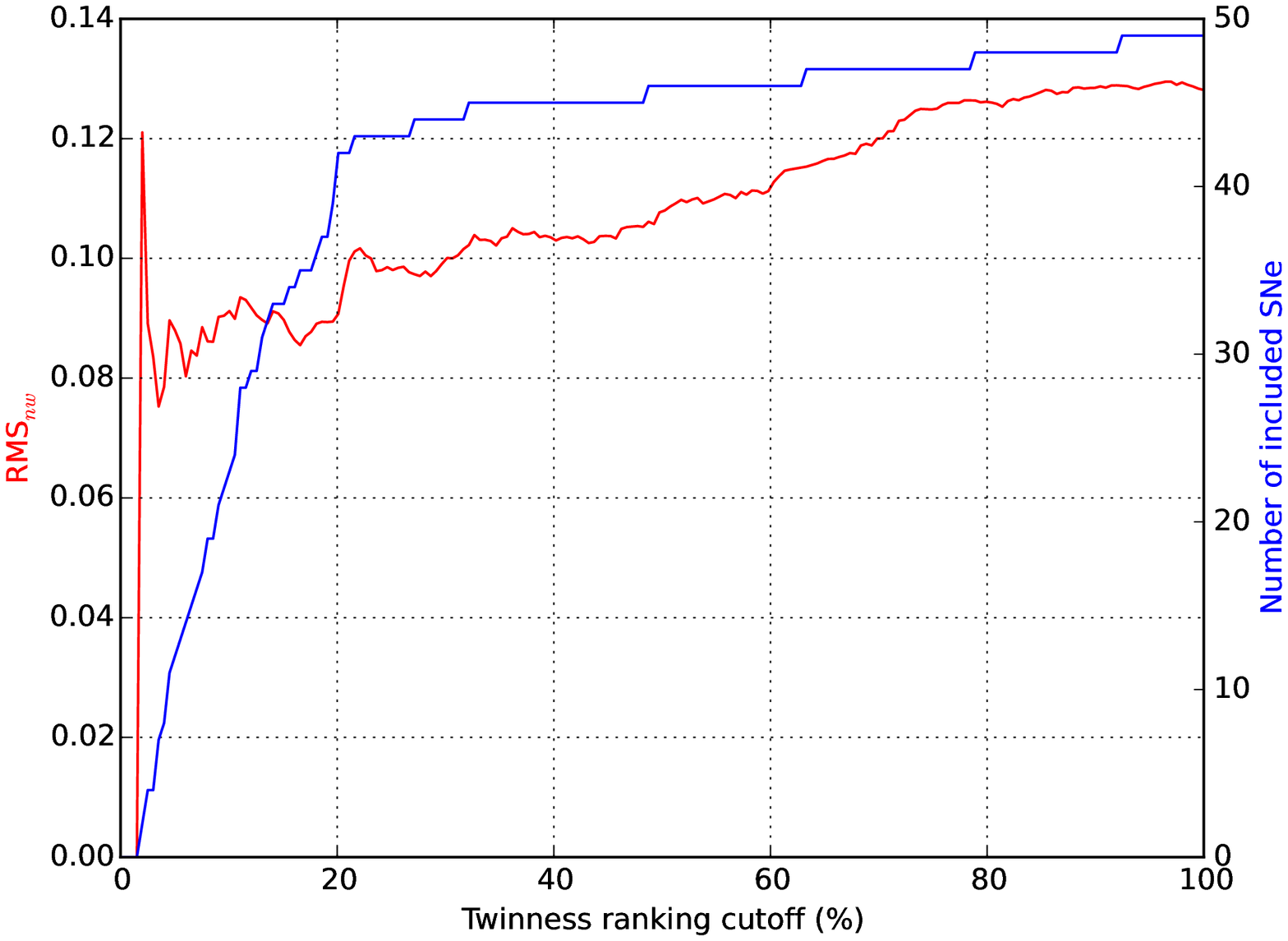}
    }
    \subfigure{
        \includegraphics[width=0.49\textwidth]{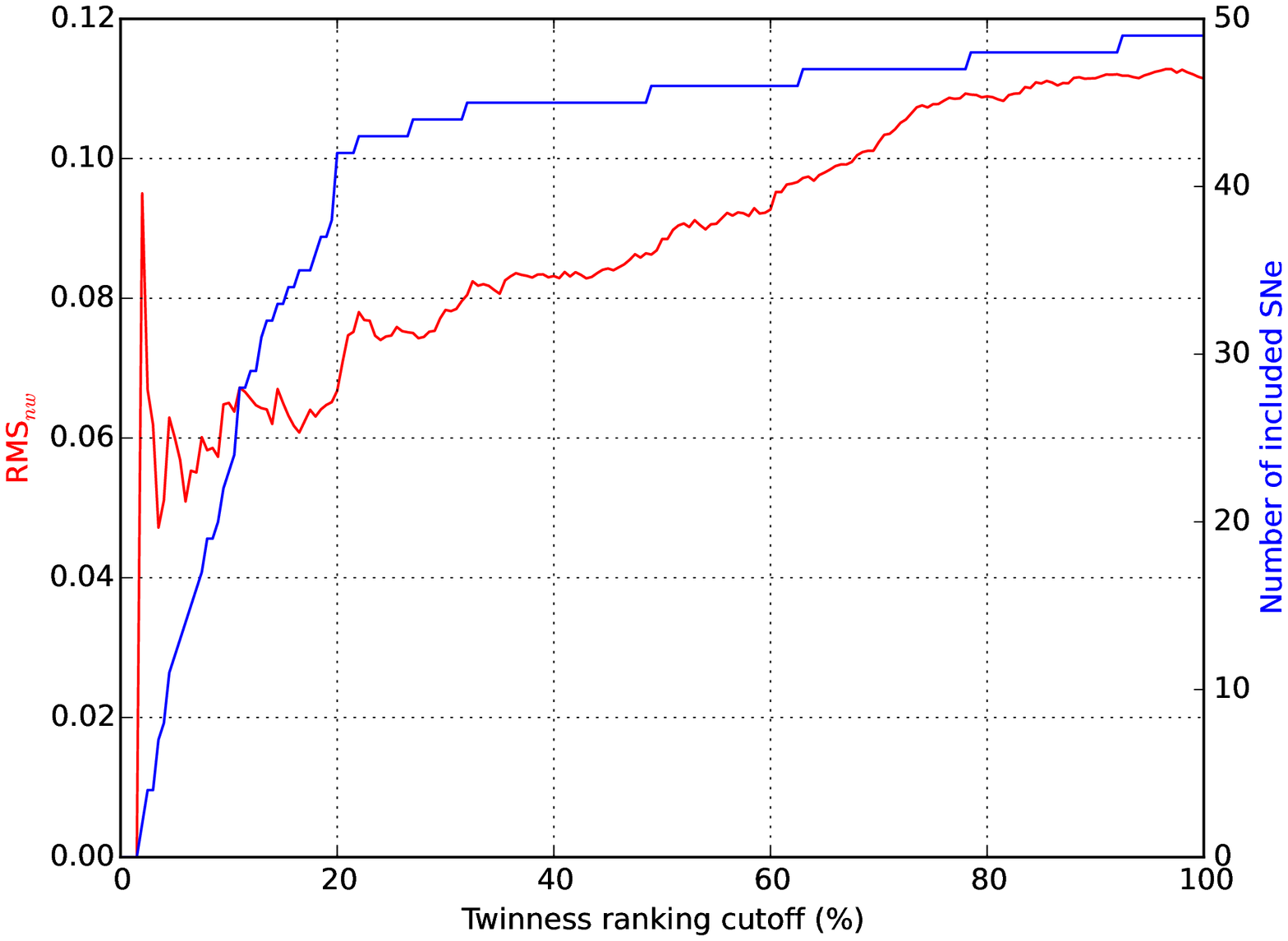}
    }
    \caption{Number-weighted root mean square (RMS$_{nw}$) as a function of twinness rank
        ($\xi$) cutoff for a realistic, single-supernova distance determination.
        The red line shows the RMS$_{nw}$, and the blue line shows the number of
        included SNe. Left: Near-maximum, $R_V=3.1$ analysis. Right: Same
        analysis with peculiar velocity and finite reference sample corrections
        applied.}
    \label{fig:wrms}
\end{figure}

To apply this to our sample, we find all the supernovae that have 
at least four pairs below some cutoff, $\xi_{\rm{cutoff}}$, (that 
is, at least 
four good twins) and we calculate the number-weighted root-mean-square 
(RMS$_{nw}$) of the distance moduli $\phi_i(\xi_{\rm{cutoff}})$ as
our measure of spread. At least four pairs are
required so that their $\phi_i$ values are reasonably well defined; 
we tested that the results are similar with a requirement of at 
least three or at least five pairs. We calculate this RMS$_{nw}$ as a 
function of $\xi_{\rm{cutoff}}$: we let in worse and worse-twinning 
pairs to the calculation of the $\phi_i$ values. We expect the
RMS$_{nw}$($\xi$) to increase with $\xi_{\rm{cutoff}}$, as more and more 
non-twin pairs are included. To summarize, the RMS$_{nw}$ formula is:
\begin{equation}
    \text{RMS$_{nw}$}(\xi_{\rm{cutoff}}) = \sqrt{\frac{\sum_i \left(\phi_i(\xi_{\rm{cutoff}})^2 (\sum_{j,
        ~\xi_{ij} < \xi_{\rm{cutoff}}} 1)\right)}{\sum_{i, ~j, ~\xi_{ij} <
        \xi_{\rm{cutoff}}}1}}
        \label{eq:rmsnw}
\end{equation}
The left panel of Figure~\ref{fig:wrms} shows the RMS$_{nw}$ as a function of $\xi$
rank for the near-maximum, $R_V=3.1$ analysis. Setting a cutoff at a $\xi$ rank of 20\% as an example, we arrive at a dispersion in the $\phi_i$ values 
of 0.091~mag, with 85\% of the supernovae included. This result is tabulated in
the fifth row of Table~\ref{tab:dispersion}. The other 15\% of 
SNe~Ia that did not have good twins could still be included albeit with a larger uncertainty. With a larger low-redshift reference sample, we
would expect that more of these supernovae would find enough pairs to be
included in our sample, and we lose very few of these. Nevertheless, for
cosmology purposes, losing part of the sample is acceptable since
the low dispersion means that the SNe that are included have more value. Additionally, there is less chance of systematic error due to an evolutionary drift in SN population properties.

\subsection{Improvements with Future Samples}\label{sec:refsize}

At high redshifts, the effects of peculiar velocity would be negligible, however, this current application analysis is affected by peculiar velocity since we are comparing low redshift supernovae to each other. As described 
in the previous section, and shown in the right two columns of Table~\ref{tab:dispersion}, the result with negligible peculiar velocity uncertainties can be estimated by assuming that each SN is affected by 
equal and independent amounts, here taken as $\Delta v =300$~km/s, and removing this contribution from the RMS$_{nw}$.
Note that this analysis is only for the purpose of estimating the likely dispersion for high-redshift SNe; we assume there will always be a 
low-redshift sample for which uncertainties due to peculiar 
velocities are relevant, and therefore we do not remove this part of the dispersion.
However, for a large low-redshift reference sample, 
the uncertainty due to random peculiar velocities would be small, approaching 
the amplitude of any coherent uncorrected bulk flow spanning 
the sample volume \citep{hui06}.

Currently there is an additional penalty due to the fact that 
a limited number of supernovae were available for comparison at the time 
the sample for this study was frozen.
Propagating the error on the $\Delta M$ values,
and assuming that even for twins there remains a finite intrinsic 
dispersion, $\sigma$, that is the same for every supernova,
the uncertainty on $\phi_i$ is given by:
\begin{equation}
    \sigma_{\phi_i} = \sigma \sqrt{1 + \frac1{\sqrt{N_{used}}}}
    \label{eqn:finite_twins}
\end{equation}
where $N_{used}$ is the number of pairs used in the estimate 
of $\phi_i$, and the RMS$_{nw}$ in Equation \ref{eq:rmsnw} is an estimator of this $\sigma_{\phi_i}$. Hence we can estimate the result that would be 
obtained with a large sample, i.e., where the number of pairs, 
$N_{used}$, is not a limitation, by dividing by the term on 
the right in Equation~\ref{eqn:finite_twins}. These results
are shown in the last line of Table~\ref{tab:dispersion}.

For such a large sample, and also
removing the peculiar velocity dispersion, we obtain the 
result shown in the right panel of Figure~\ref{fig:wrms}. 
Here, with a cut on twinness rank at 20\% and using $R_V=3.1$,
we would expect $\sigma_{\phi_i} \sim 0.067$~mag with 85\%
of the sample included. For the $R_V=2.8$ analysis, 
we would expect $\sigma_{\phi_i} \sim 0.057$~mag. The simple 
approach of Section~\ref{sec:highz} for realistically
implementing the twin method thus, in the limit of a 
large sample size, gives results consistent with the RMS 
results shown for the pair-wise analysis of 
Section~\ref{sec:deltam}.

Future use of the $\sim$200--300~SNe~Ia anticipated in 
the final SNfactory dataset as a reference 
is expected to gain much of this large-sample dispersion improvement.
The value of $N_{used}$ 
in Equation~\ref{eqn:finite_twins} will scale up in proportion to the
overall sample size. Therefore, we forecast that this sample 
of $\sim$200--300~SNe~Ia already will be sufficient to reduce 
the dispersion to $\sigma_{\phi_i} \sim 0.073$~mag. 
For $R_V=2.8$, $\sigma_{\phi_i} \sim 0.062$~mag
would be expected.

Enlarging the sample should also improve both the quality of the
best twins and the number of supernovae that are able to find twins.
Empirically, we found a 20\% decrease in the dispersion when the
training sample was combined with the validation sample. The finite
reference sample correction only predicts a 6\% decrease, so the
majority of this improvement is likely due to the supernovae in
the full sample having better twins available. In the most optimistic
scenario, where there is no dispersion in brightness between 
supernovae that are truly twins, the relative brightness 
uncertainty will be entirely limited by how well matched  
the best twins available in the reference sample are for a 
new supernova. We note that we observe this extra decrease in dispersion,
but this improvement depends heavily on the properties of any
subpopulations of SNe Ia, so we cannot predict how rapidly the
dispersion will improve as the sample size grows.
Future studies with larger reference samples will be able
to measure the true dispersion of the best twins, and determine whether
it varies for subpopulations of SNe Ia.

The twins method has obvious cosmological applications nearby, such as measuring
distances to nearby galaxies and mapping the peculiar velocity field. In future
work, we will explore the performance characteristics of the twins method as a
function of resolution and signal-to-noise. In the current analysis, the SNIFS
data were binned to a resolution of R $\sim$ 150. Our lowest signal-to-noise
spectra have a signal-to-noise of $\sim 12$ at this resolution and show no
signs of degradation in their brightness dispersion for best twins. These data
characteristics are modest, suggesting that the twins technique is likely to be
within reach of experiments such as JWST and WFIRST, even at high redshifts.

\subsection{Phase Determination from Lightcurve}\label{sec:otherstuff}

An important implementation consideration is that the twins analysis
requires knowledge of the phase of a spectrum when comparing it to the
reference sample. Currently, this is done using a SALT2.4 fit to a full 
lightcurve. An experiment that applied the near-maximum version of the 
twins method 
would need to obtain one good spectrum near maximum along with crude 
photometry at other points on the lightcurve sufficient to provide 
the date of maximum. It may be possible to estimate the date of 
maximum directly from the twins method, along
the lines of existing SN typing codes, but this has
not yet been attempted and is left for future work.

The uncertainty of the phase determination will propagate into the final
uncertainty on the relative distance. For our sample, the median uncertainty on
the day of maximum is 0.28 days, which means that we have an uncertainty of
0.4 days on our phases when comparing two SNe. We redid our analysis with
Gaussian Process predictions offset from the measured maximum phase to get an
estimate of the contribution to the brightness dispersion from phase errors.
For a 0.4 day offset, we get an additional dispersion of 0.03~mag, while for a
1 day offset we get an additional dispersion of 0.05~mag. Note that these
values are upper limits on the contributions to the dispersion since we have
introduced new phase errors in addition to the ones already present in the
data. Regardless, this uncertainty is still significantly smaller than even the
smallest dispersion listed in Table~\ref{tab:dispersion}, so the phase error
cannot be the main contributor. Future experiments
will apparently need to constrain the day of maximum of SNe to an accuracy
comparable to that of the current analysis if they are to obtain comparably
tight magnitude dispersions.

\section{Conclusions}\label{sec:conclusions}

We have shown that Type Ia supernovae that are 
spectroscopic twins have a lower luminosity dispersion 
than spectroscopically dissimilar Type Ia supernovae. 
Rather than using the lightcurves to standardize 
the supernova samples, future surveys could standardize supernovae 
employing a single spectrum measured near maximum 
and applying the twins method. With the subset of 
SNfactory data used in this paper, the twins method can
standardize supernovae to $\sim$0.08~mag at low redshifts. After removing uncertainties due to host-galaxy peculiar velocities the resulting dispersion would be 
$\sim$0.07~mag. This is the value expected at high 
redshifts, where the fractional contribution due to 
peculiar velocities is negligible. It is estimated that 
the twins method may be able standardize high-redshift 
supernovae with a dispersion of 0.06--0.07~mag once a
larger sample 
becomes available as a reference.
The twins dispersions are 
as good as the best known standardization methods to 
date. While implicit in the concept of twin SNe~Ia, 
future work, requiring a larger sample, will be needed 
to study the extent to which the non-parametric twins 
method subsumes other parametric standardization methods.

An important question in supernova cosmology studies is what portion of the
observed distance dispersion is due to astrophysical differences among the
supernovae that are not accounted for by current analysis techniques. Our
results imply that at least 3/4 of the variance in Hubble residuals from a
SALT-like lightcurve-based standardization is due to such differences among the
supernovae. That is, the SN astrophysical variation dominates over other common
concerns, such as variation in dust extinction laws and uncorrelated
calibration errors.

Our analysis shows that spectrophotometric data of supernovae is 
extremely valuable, and we find a low dispersion even if only a 
single spectrum is taken near maximum. This 
standardization method also has the advantage that it does not 
require rest-frame infra-red data, so it can be used for supernovae 
at much higher redshifts than is possible with standardization that 
requires such data, given current and planned near-IR observing 
capabilities.

The fact that the magnitude dispersion dramatically improves for better
twins --- whether or not the magnitudes are standardized using conventional
methods --- implies that there are subpopulations of Type Ia SNe with
different mean absolute magnitudes that can be recognized by their spectra.
The relative frequency of these different SNe~Ia affects the mean brightness
of an ensemble measured at a given redshift. Thus, if the demographics
of the SNe~Ia changes with redshift \citep[e.g.][]{rigault13,childress14,rigault15},
current measurements of the cosmological parameters could be biased.
The twins analysis is, in principle, capable of handling any drift in the
demographics of the SN~Ia population with redshift without any 
modifications.  If all major subpopulations are well represented in 
the reference sample, the twins method will be comparing new SNe to 
reference SNe within their subpopulations. The method is therefore 
limited only by the inherent dispersion of the subpopulations. Unusual 
SNe, or previously unobserved subpopulations of SNe, can be
identified and rejected automatically since they will not twin well 
with any SNe in the reference set.

Further work is needed to determine the minimal spectral information 
required to achieve these twin results. In addition, while 
weighting by the covariance between different phases has been explored here, it is possible that weighting by the covariance between both phases and wavelengths might produce
an even smaller dispersion. The final SNfactory dataset will
be much larger, allowing such a broader range of studies. The twins 
method could also be applied to analyses of supernova subtypes 
beyond those pioneered by \citet{branch06}. The twinness rank, 
$\xi$, provides a natural way to quantitatively identify subtypes 
of Type~Ia supernovae based on groups of closely matched SNe.

\acknowledgements

We thank Dan Birchall for observing assistance, the technical and
scientific staffs of the Palomar Observatory, the High Performance
Wireless Radio Network (HPWREN), and the University of Hawaii 2.2~m
telescope.  We recognize the significant cultural role of Mauna Kea
within the indigenous Hawaiian community, and we appreciate the
opportunity to conduct observations from this revered site.  This
work was supported in part by the Director, Office of Science,
Office of High Energy Physics, of the U.S. Department of Energy
under Contract No. DE-AC02- 05CH11231.  Support in France was
provided by CNRS/IN2P3, CNRS/INSU, and PNC; LPNHE acknowledges
support from LABEX ILP, supported by French state funds managed by
the ANR within the Investissements d'Avenir programme under reference
ANR-11-IDEX-0004-02.  NC is grateful to the LABEX Lyon Institute
of Origins (ANR-10-LABX-0066) of the Universit\'e de Lyon for its
financial support within the program ``Investissements d'Avenir''
(ANR-11-IDEX-0007) of the French government operated by the National
Research Agency (ANR).  Support in Germany was provided by the DFG
through TRR33 ``The Dark Universe;'' and in China from Tsinghua
University 985 grant and NSFC grant No~11173017.  Some results were
obtained using resources and support from the National Energy
Research Scientific Computing Center, supported by the Director,
Office of Science, Office of Advanced Scientific Computing Research,
of the U.S. Department of Energy under Contract No. DE-AC02-05CH11231.
HPWREN is funded by National Science Foundation Grant Number
ANI-0087344, and the University of California, San Diego.

\bibliographystyle{apj}

\appendix
\section{Gaussian Process Regression}\label{sec:gp_params}

We train a Gaussian Process to perform interpolation on our data with the following parameters.
For the mean function, we choose the spectral template of 
\citet{hsiao} and allow one hyperparameter, $\theta_a$, to handle
the overall flux normalization. 
Note that the mean function is what the GP prediction will default to in the absence
of coverage by training data. For our kernel, we use a squared exponential of the form:

\begin{equation}
k
\begin{pmatrix}
\begin{bmatrix}
\mathrm{ln}~\lambda_i \\
t_i
\end{bmatrix}
,
\begin{bmatrix}
\mathrm{ln}~\lambda_j \\
t_j
\end{bmatrix}
\end{pmatrix}
= \theta_s^2 \mathrm{exp}
\begin{pmatrix}
-\frac{1}{2}
\begin{bmatrix}
\mathrm{ln}~\lambda_i - \mathrm{ln}~\lambda_j \\
t_i - t_j
\end{bmatrix}
^T
\begin{bmatrix}
1/\theta_\lambda^2 & 0 \\
0 & 1/\theta_t^2
\end{bmatrix}
\begin{bmatrix}
\mathrm{ln}~\lambda_i - \mathrm{ln}~\lambda_j \\
t_i - t_j
\end{bmatrix}
\end{pmatrix}
+ \theta^2_n\delta_{t_i,t_j}
\end{equation}

We use the difference in $\mathrm{ln}~\lambda$ as opposed to $\lambda$ since velocity space is more physically
relevant to supernova explosions. $\theta_s$ is the amplitude hyperparameter and gives the scale of the GP prediction
error when sufficiently far away from training data.  $\theta_\lambda$ and $\theta_t$ are, respectively, the length
scales in log wavelength and time.  $\theta_n$ is the time nugget, which differs from noise in that it 
is non-zero only when two data points are at the same phase, and accounts for wavelength-independent calibration
errors and noise.

\begin{figure}[ht]
\centering
\includegraphics[scale=0.45]{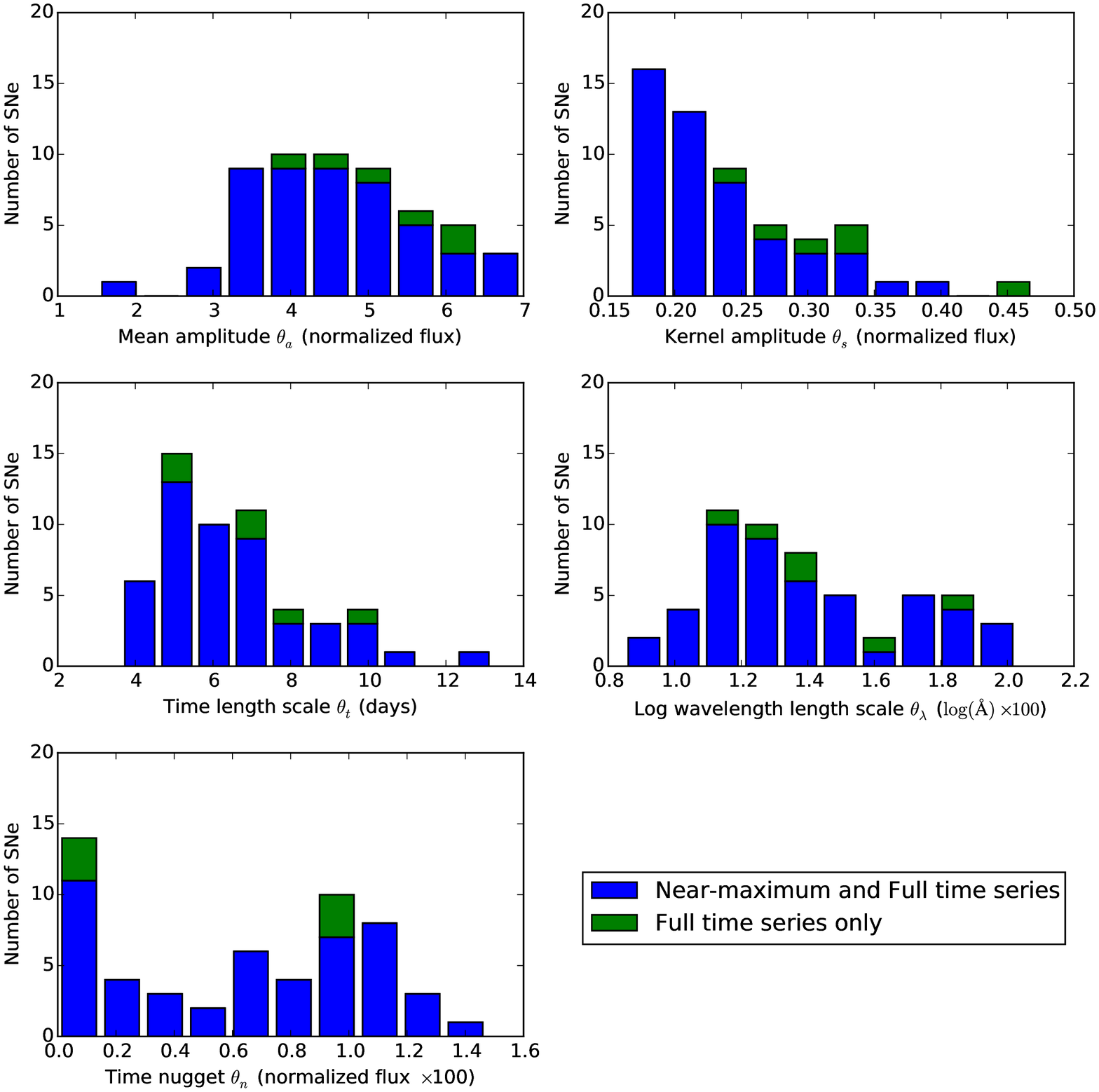}
\caption{Histograms of the various hyperparameters for the supernovae in both the near-maximum and full time series
    samples. Top left to bottom left are: mean amplitude ($\theta_a$), kernel amplitude ($\theta_s$),
    time length scale ($\theta_t$), log wavelength length scale ($\theta_\lambda$), and time nugget ($\theta_n$)}
\label{hparams}
\end{figure}

We use maximum likelihood estimation for each individual time series to calculate the hyperparameters for the mean
and the kernel simultaneously. We use Minuit2 as our optimizer \citep{james75}. We are then able to use these optimized 
hyperparameters to make predictions at any desired wavelengths and phases. Figure~\ref{hparams} show histograms of
the hyperparameter values for each of the five hyperparameters used in this analysis.  The range of values for
each of these hyperparameters is reasonable. 

The supernovae have been shifted to a common redshift, so we expect $\theta_a$
to have a similar dispersion as uncorrected supernovae in flux space, and
that is indeed what we see. For the time length scale, $\theta_t$, we have 
values ranging from $\sim$4 to $\sim$12~nights.  Although SNe~Ia vary 
on few day timescales (as is evidenced by examination of their spectra), this hyperparameter value is affected by the data sampling as well as the
overall smoothness of the data since a shorter time length scale will allow for more variation in the
GP predictions.  The range of $\theta_\lambda$ corresponds to velocities between $\sim$2250~km~s$^{-1}$ and $\sim$7750~km~s$^{-1}$.
Or, in terms of wavelength, $\sim$25\AA-85\AA~at the blue end and 
$\sim$65\AA-215\AA~at the red end. This is a reasonable range given that 
SN feature widths are typically several thousand km~s$^{-1}$.

\section{Phase Weighting Covariance Matrix}
\label{sec:weightcov}

For the full spectral time series analysis in Section \ref{sec:fullts},
the correlation in brightness dispersion between any two phases, $C(p_i, p_k)$ needs to be determined. We estimate this function
from the data. This is done by performing a calculation as in the near-maximum study centered around all
of $-5, 0, 5, 10, 15, 20$ and $25$ days after maximum.  The
correlation matrix of these data is calculated using the subset of pairs that is in each of
the phase bins. It is worth noting that different pairs of supernovae can be
missing in different samples depending on the cadence of their observations.
Hence not all of the same pairs are used for each entry in this correlation
matrix, making it very poorly conditioned.

In order to be useful the correlation function must 
work given any two phases $p_i$ and $p_k$, not just 
two chosen from the samples above. To regularize the 
computation, we assume that the correlation between 
two phases is given by:
\
\begin{equation}
    D(p_i, p_k) = \exp(-\alpha |p_i-p_k|)
\end{equation}

and we then solve for $\alpha$ by fitting to the 
results from the above sample. We obtain 
$\alpha = 0.053\text{~day}^{-1}$ for $R_V=3.1$ and 
$\alpha = 0.060\text{~day}^{-1}$ for $R_V=2.8$. Note 
that this indicates a very high correlation between 
nearby phases, so only limited information is gained
from adding additional nearby phases. Interestingly, $\alpha^{-1} \sim 16$ days. This is
comparable to the 15 day difference in brightness $\Delta M_{15}$ that is 
commonly used to standardize supernovae \citep{phillips93}.

We are interested in optimizing the weighting for 
good twins rather than for the sample as a whole. 
Hence, when calculating the variance, we apply the
near-maximum twinning algorithm for each sample 
independently, and we take the variance of the best 
10\% of twins in each sample. Ideally we would only 
use the best 10\% as well when calculating the 
full covariance matrix, but our sample size is currently
too small to make this statistically feasible since 
overlapping pairs are required when calculating the 
correlation between two bins. A spline is then fit 
through the calculated variances in order to interpolate 
to any phase. Combining the correlation and variances, 
we can calculate the covariance matrix $C(p_i, p_k)$ for
any combination of phases. 

\section{Monte Carlo simulation}
\label{sec:monte_carlo}

We used a Monte Carlo simulation to calculate the covariance of the RMS bins.
The Monte Carlo simulation involves the following steps:

\begin{enumerate}
    \item \label{item:magpdf} Generate a probability density function (PDF) of the distribution of
        magnitudes.
    \item \label{item:twindist} Generate a PDF of the distribution of twinness
        for each value of $\Delta M$.
    \item \label{item:toystart} Randomly choose a set of magnitudes from the
        magnitude distribution to generate a mock sample.
    \item \label{item:twinsample} Using the same pairs as in our sample, assign a twinness to each value
        of $\Delta M$ using the distribution from Step \ref{item:twindist}
    \item \label{item:toyend} Calculate the twinness ranks, bin the twins
        accordingly and calculate the RMS $\Delta M$ for each group.
    \item Repeat Steps \ref{item:toystart} to \ref{item:toyend} 1000 times.
    \item Calculate the covariance matrix from the generated data.
\end{enumerate}

Steps \ref{item:magpdf} and \ref{item:twindist} were done by estimating PDFs from the data using Kernel
Density Estimators (KDEs). This involves generating the PDF from the input data
by effectively convolving the input data with some sort of a kernel. We use a
Gaussian kernel, so the PDF of magnitudes in Step \ref{item:magpdf} is given by:
\begin{equation}
    \hat f(M) = \frac1n \sum_{i=1}^{\#SN} \frac1{\sqrt{2 \pi \sigma_M^2}}
    \exp\left(-\frac{(M-M_i)^2}{2 \sigma_M^2}\right)
\end{equation}
where the $M_i$ are the given magnitudes of supernovae. Note that $\sigma_M$ is
a free parameter that must be chosen. Similarly, for the PDF of twinness as a
function of $\Delta M$ in Step \ref{item:twindist}, we follow a similar procedure in two dimensions:
\begin{equation}
    \hat g(t | \Delta M) = \frac1n \sum_{i=1}^{\#pairs} \frac1{2 \pi
\sigma_{\Delta M} \sigma_t} \exp \left(-\frac{(\Delta M-\Delta M_i)^2}{2
\sigma_{\Delta M}^2} -\frac{(t-t_i)^2}{2 \sigma_t^2}\right)
\end{equation}
where $\sigma_{\Delta m}$ and $\sigma_t$ are free parameters that must be
chosen. We chose the kernel parameters by hand by taking the lowest values that
gave smooth PDFs. This resulted in $\sigma_M = 0.02$ and $\sigma_t = 0.2
\times\text{median(twinness)}$ since the twinness changes in scale depending on
the formula chosen. These were checked for both the near-maximum and weighted
full spectral timeseries analyses, and gave good results in both cases. $\sigma_{\Delta
M}$ is then given by $\sqrt{2}\sigma_M$ by error propagation. The results of the
Monte Carlo were not used to tune these parameters, purely the PDF smoothness.

In the real data, some supernovae find twins much more often than others. This
is thought to be due to the fact that if supernovae are close to the ``mean''
supernova, then they are in general good twins with most fairly normal
supernovae. There are also several supernovae that do not twin well at all. In
order to handle these in the Monte Carlo, a ``tilt'' was added to each simulated
pair. The tilt is added to the twinness of each pair that the supernova appears
in, so that the final twinness is of the pair is given by:
\begin{equation}
    t_{ij} = t_{\Delta M_{ij}} + tilt_i + tilt_j
\end{equation}
where $t_{\Delta M_{ij}}$ is given by the PDFs described above. The tilt for
each supernova is sampled from a PDF of a Gaussian of width $0.75~\sigma_t$
centered at 0 added to a uniform distribution between twinness values of 0 and
$10~\sigma_t$. The PDF is weighted so that 90\% of the weight goes to the
Gaussian and 10\% goes to the flat line. The Gaussian makes some supernovae be
slightly over-represented in the lowest twinness bins. The flat line allows for
supernovae that do not twin well with anything. These numbers were chosen by
matching the square sum of bin multiplicities in the simulations to the real
data (a straightforward calculation shows that this value is what the variance
is scaled by, assuming an underlying Gaussian distribution in magnitudes). The
results are not sensitive to small changes in these values.

We generate mock datasets from these PDFs (Steps \ref{item:toystart} through
\ref{item:toyend}) using inverse transform sampling. The
mock datasets from this simulation are consistent with the actual twins result.
Running 1000 mocks, we get the covariance matrix for the $R_V=3.1$ near-maximum sample. The mean RMS of the
mocks comes out at 0.1253 with a standard deviation of 0.0206, which is
consistent with the value of 0.1281 for the real data. As a cross check, the covariance matrix was estimated analytically assuming no twinness 
and an
underlying Gaussian distribution which gave consistent, but less conservative,
results. The same procedure was run for each dataset to generate individual covariance
matrices.

\section{Bin occupancies}

\label{sec:bin_occupancies}

The bin occupancies for the near-maximum analysis with $R_V=3.1$
are shown in Table
\ref{tab:occupancy}. This table illustrates the fact that there are some SNe
that twin far better than others. There are also several SNe that are not
good twins with any SNe, and they are automatically rejected or deweighted
from the analyses as a result.

\begin{table}[ht]
    \caption{Occupancy of SNe in twinness bins for the near-maximum $R_V=3.1$
    analysis as shown in Figure~\ref{fig:f3050}. The bin labeled 0 contains the best twins, and the bin
    labeled 16 contains the worst twins.}
    \label{tab:occupancy}
    \tiny
    \centering
    \begin{tabular}{lrrrrrrrrrrrrrrrrr}
\toprule
{} &  0  &  1  &  2  &  3  &  4  &  5  &  6  &  7  &  8  &  9  &  10 &  11 &  12 &  13 &  14 &  15 &  16 \\
\midrule
SNF20050728-006 &   3 &   2 &   8 &   4 &   8 &   4 &   2 &   2 &   2 &   2 &   1 &   2 &   2 &   1 &   2 &   0 &   0 \\
SNF20060511-014 &   1 &   3 &   1 &   0 &   1 &   3 &   0 &   2 &   1 &   5 &   1 &   4 &   8 &   1 &   2 &   7 &   3 \\
SNF20060512-001 &   1 &   1 &   1 &   1 &   0 &   0 &   0 &   0 &   2 &   3 &   1 &   0 &   3 &   8 &   6 &   6 &  13 \\
SNF20060526-003 &   6 &   4 &   2 &   2 &   5 &   6 &   6 &   2 &   4 &   1 &   1 &   1 &   2 &   1 &   0 &   1 &   0 \\
SNF20050624-000 &   1 &   1 &   1 &   1 &   3 &   1 &   0 &   1 &   2 &   1 &   2 &   0 &   0 &   1 &   0 &   0 &   0 \\
SNF20060621-015 &   5 &   3 &   4 &   5 &   3 &   5 &   6 &   3 &   2 &   3 &   1 &   2 &   0 &   1 &   1 &   1 &   1 \\
SNF20060907-000 &   4 &   1 &   1 &   6 &   3 &   9 &   2 &   3 &   4 &   2 &   2 &   2 &   1 &   1 &   3 &   1 &   2 \\
SNF20060908-004 &   1 &   0 &   2 &   2 &   3 &   3 &   2 &   1 &   2 &   4 &   3 &   3 &   2 &   3 &   0 &   0 &   0 \\
SNF20070330-024 &   0 &   0 &   2 &   4 &   2 &   0 &   1 &   5 &   1 &   4 &   3 &   4 &   4 &   5 &   4 &   3 &   1 \\
SNF20070403-001 &   0 &   4 &   2 &   3 &   2 &   5 &   6 &   3 &   3 &   1 &   2 &   4 &   1 &   3 &   3 &   2 &   0 \\
SNF20070506-006 &   0 &   4 &   4 &   2 &   0 &   4 &   3 &   0 &   2 &   2 &   2 &   2 &   3 &   4 &   7 &   6 &   3 \\
SNF20070531-011 &   3 &   5 &   3 &   2 &   3 &   1 &   2 &   4 &   1 &   2 &   4 &   1 &   3 &   1 &   2 &   1 &   4 \\
SNF20070701-005 &   1 &   1 &   2 &   2 &   1 &   2 &   1 &   1 &   2 &   5 &   6 &   5 &   6 &   2 &   4 &   0 &   3 \\
SNF20070725-001 &   6 &   5 &   3 &   2 &   3 &   4 &   2 &   6 &   5 &   1 &   1 &   1 &   0 &   0 &   0 &   1 &   0 \\
SNF20070712-003 &   3 &   0 &   0 &   1 &   2 &   0 &   7 &   2 &   3 &   1 &   4 &   0 &   2 &   3 &   0 &   3 &   2 \\
SNF20070727-016 &   1 &   0 &   2 &   1 &   2 &   0 &   2 &   1 &   1 &   2 &   1 &   4 &   3 &   4 &   3 &   7 &   6 \\
SNF20070802-000 &   0 &   0 &   1 &   0 &   0 &   0 &   1 &   1 &   4 &   1 &   1 &   6 &   2 &   8 &   4 &   8 &  10 \\
SNF20070806-026 &   3 &   2 &   2 &   2 &   5 &   4 &   3 &   2 &   0 &   1 &   2 &   3 &   3 &   3 &   3 &   0 &   1 \\
SNF20070817-003 &   0 &   0 &   2 &   4 &   1 &   2 &   1 &   5 &   6 &   3 &   4 &   2 &   5 &   2 &   3 &   2 &   3 \\
SNF20070818-001 &   0 &   0 &   1 &   0 &   1 &   2 &   3 &   3 &   5 &   2 &   3 &   1 &   6 &   3 &   3 &   3 &   2 \\
SNF20070820-000 &   3 &   3 &   5 &   5 &   4 &   5 &   0 &   1 &   2 &   3 &   2 &   3 &   2 &   2 &   0 &   1 &   0 \\
SNF20070831-015 &   1 &   2 &   2 &   2 &   2 &   2 &   0 &   3 &   3 &   3 &   4 &   6 &   3 &   4 &   6 &   1 &   2 \\
SNF20071003-016 &   3 &   6 &   1 &   5 &   1 &   2 &   3 &   2 &   2 &   1 &   3 &   2 &   3 &   0 &   3 &   1 &   0 \\
SNF20070902-018 &   1 &   2 &   0 &   4 &   4 &   2 &   3 &   1 &   0 &   1 &   1 &   0 &   5 &   1 &   2 &   2 &   0 \\
SNF20071015-000 &   6 &   3 &   3 &   3 &   3 &   1 &   0 &   3 &   3 &   7 &   3 &   1 &   2 &   2 &   0 &   1 &   1 \\
SNF20080507-000 &   2 &   0 &   3 &   2 &   2 &   1 &   3 &   2 &   1 &   4 &   7 &   7 &   4 &   4 &   1 &   2 &   1 \\
SNF20080510-001 &   2 &   4 &   5 &   1 &   3 &   6 &   3 &   2 &   2 &   2 &   4 &   4 &   0 &   4 &   2 &   2 &   1 \\
SNF20080512-010 &   2 &   7 &   3 &   4 &   2 &   5 &   5 &   4 &   3 &   3 &   2 &   3 &   1 &   1 &   0 &   1 &   1 \\
SNF20080516-000 &   0 &   5 &   6 &   4 &   2 &   2 &   1 &   3 &   2 &   2 &   2 &   5 &   3 &   2 &   1 &   0 &   0 \\
SNF20080531-000 &   6 &   4 &   1 &   4 &   3 &   6 &   2 &   5 &   1 &   0 &   0 &   3 &   2 &   1 &   2 &   1 &   2 \\
SNF20080610-000 &   2 &   2 &   2 &   3 &   2 &   4 &   2 &   3 &   2 &   5 &   5 &   2 &   2 &   0 &   7 &   2 &   1 \\
SNF20080620-000 &   7 &   5 &   2 &   1 &   3 &   0 &   2 &   3 &   4 &   2 &   2 &   1 &   1 &   3 &   3 &   4 &   3 \\
SNF20080623-001 &   2 &   2 &   5 &   2 &   0 &   1 &   6 &   5 &   3 &   7 &   2 &   2 &   3 &   1 &   1 &   3 &   3 \\
SNF20080714-008 &   0 &   0 &   0 &   0 &   0 &   0 &   1 &   1 &   0 &   0 &   0 &   0 &   0 &   6 &   4 &  10 &  14 \\
SNF20080725-004 &   1 &   2 &   3 &   3 &   3 &   2 &   0 &   3 &   2 &   3 &   3 &   1 &   2 &   3 &   1 &   0 &   1 \\
SNF20080803-000 &   4 &   5 &   9 &   2 &   5 &   4 &   3 &   3 &   3 &   2 &   1 &   3 &   0 &   2 &   0 &   1 &   0 \\
SNF20080810-001 &   2 &   1 &   2 &   1 &   1 &   3 &   2 &   3 &   2 &   4 &   5 &   4 &   3 &   3 &   6 &   2 &   4 \\
SNF20080821-000 &   6 &   5 &   2 &   3 &   5 &   4 &   8 &   4 &   1 &   3 &   2 &   0 &   1 &   2 &   0 &   1 &   0 \\
SNF20080822-005 &   3 &   3 &   1 &   0 &   1 &   1 &   1 &   4 &   2 &   2 &   2 &   3 &   3 &   3 &   3 &   4 &   1 \\
SNF20080825-010 &   4 &   5 &   2 &   8 &   3 &   2 &   3 &   1 &   1 &   1 &   2 &   2 &   4 &   3 &   4 &   0 &   2 \\
SNF20080909-030 &   0 &   1 &   0 &   0 &   0 &   0 &   0 &   0 &   0 &   0 &   0 &   0 &   0 &   2 &   0 &   1 &   0 \\
SN2005hc        &   1 &   2 &   2 &   1 &   1 &   1 &   3 &   1 &   3 &   3 &   4 &   3 &   5 &   1 &   4 &   7 &   2 \\
SN2006cj        &   7 &   1 &   4 &   2 &   3 &   1 &   5 &   4 &   3 &   6 &   5 &   0 &   1 &   1 &   2 &   1 &   1 \\
SN2007bd        &   0 &   3 &   2 &   2 &   4 &   0 &   1 &   3 &   6 &   2 &   2 &   3 &   3 &   3 &   5 &   2 &   1 \\
SN2007nq        &   4 &   1 &   2 &   1 &   5 &   1 &   3 &   1 &   2 &   2 &   3 &   3 &   1 &   2 &   1 &   2 &   0 \\
PTF09dlc        &   2 &   0 &   1 &   0 &   1 &   2 &   1 &   2 &   4 &   1 &   1 &   4 &   6 &   2 &   5 &   6 &   4 \\
PTF09dnp        &   0 &   0 &   0 &   0 &   0 &   0 &   0 &   1 &   0 &   1 &   2 &   1 &   1 &   3 &   5 &   5 &  18 \\
PTF09fox        &   7 &   5 &   5 &   4 &   4 &   3 &   2 &   2 &   2 &   2 &   3 &   3 &   2 &   0 &   1 &   2 &   0 \\
PTF09foz        &   1 &   3 &   3 &   7 &   5 &   2 &   5 &   3 &   7 &   2 &   1 &   2 &   1 &   2 &   1 &   1 &   1 \\
\bottomrule
\end{tabular}
\end{table}

\section{Full twin pairing data}

\label{sec:full_data}

The full twin pairing data for the near-maximum $R_V=3.1$ analysis
is shown for 20 supernovae pairs in Table \ref{tab:20_data}, ordered by the
twinness value $\xi$. The full data for all of the pairs is available in the
electronic version.

\begin{table}[ht]
    \caption{Full data for the best 20 supernovae pairs from the near-maximum
    $R_V=3.1$ analysis. The full data for all pairs is available in the
electronic version.}
    \label{tab:20_data}
    \centering
    \begin{tabular}{llccc}
\toprule
\multicolumn{1}{c}{SN$_A$} &
\multicolumn{1}{c}{SN$_B$} &  $\xi$    & $\Delta M$ &      $\Delta$E(B-V) \\
\midrule
SNF20060621-015 &  SNF20070725-001 &  0.128497 & -0.037632 &  0.00549 \\
       SN2006cj &  SNF20071015-000 &  0.183916 &  0.057913 & -0.33847 \\
SNF20070727-016 &  SNF20080822-005 &  0.190385 &  0.069850 & -0.06035 \\
SNF20070806-026 &  SNF20080825-010 &  0.220105 &  0.136250 &  0.01811 \\
SNF20080531-000 &  SNF20080623-001 &  0.230594 & -0.101350 &  0.03382 \\
SNF20060621-015 &  SNF20060907-000 &  0.231818 & -0.149275 &  0.02178 \\
       SN2006cj &  SNF20080821-000 &  0.234615 &  0.105564 & -0.04483 \\
SNF20071003-016 &  SNF20080620-000 &  0.249301 & -0.061353 &  0.05521 \\
SNF20071015-000 &  SNF20080821-000 &  0.270979 &  0.048357 &  0.29497 \\
SNF20070531-011 &  SNF20080531-000 &  0.277972 & -0.059594 &  0.01894 \\
       SN2006cj &  SNF20070725-001 &  0.282517 & -0.150393 &  0.08062 \\
       SN2007nq &  SNF20070531-011 &  0.288287 &  0.021982 &  0.00652 \\
       PTF09fox &  SNF20080821-000 &  0.294231 &  0.232054 & -0.08943 \\
       PTF09fox &  SNF20080803-000 &  0.295455 &  0.158407 & -0.15035 \\
SNF20070725-001 &  SNF20071015-000 &  0.304371 &  0.183241 & -0.41736 \\
SNF20070806-026 &  SNF20080512-010 &  0.311364 &  0.109478 & -0.00583 \\
SNF20060526-003 &  SNF20080821-000 &  0.317483 &  0.127348 & -0.03741 \\
SNF20060907-000 &  SNF20070820-000 &  0.318182 &  0.221476 & -0.20660 \\
SNF20080803-000 &  SNF20080821-000 &  0.321503 &  0.082138 &  0.05966 \\
SNF20060526-003 &  SNF20070725-001 &  0.322028 & -0.122426 &  0.09097 \\
\multicolumn{1}{c}{\nodata} &
          \multicolumn{1}{c}{\nodata} & 
          \multicolumn{1}{c}{\nodata} &
          \multicolumn{1}{c}{\nodata} & 
          \multicolumn{1}{c}{\nodata} \\
\bottomrule
\end{tabular}
\end{table}

\end{document}